# Challenges and opportunities in piezoelectric polymers: Effect of oriented amorphous fraction in ferroelectric semicrystalline polymers

Guanchun Rui[1] | Elshad Allahyarov[2,3,4] | Zhiwen Zhu[5] | Yanfei Huang[6] | Thumawadee Wongwirat[1] | Qin Zou[1] | Philip L. Taylor[2] | Lei Zhu[1]

[1]Department of Macromolecular Science and Engineering, Case Western Reserve University, Cleveland, Ohio, USA

[2]Department of Physics, Case Western Reserve University, Cleveland, Ohio, USA

[3]Institut für Theoretische Physik II: Weiche Materie, Heinrich-Heine Universität Düsseldorf, Düsseldorf, Germany

[4]Theoretical Department, Joint Institute for High Temperatures, Russian Academy of Sciences, Moscow, Russia

[5]Engineering Research Center of Novel Equipment for Polymer Processing, Guangdong Provincial Key Laboratory of Technique and Equipment for Macromolecular Advanced Manufacturing, South China University of Technology, Guangzhou, China

[6]College of Materials Science and Engineering, Shenzhen University, Shenzhen, China

**Correspondence**
Lei Zhu and Philip L. Taylor.
Email: lxz121@case.edu and plt@case.edu

**Abstract**

Despite extensive research on piezoelectric polymers since the discovery of piezoelectric poly(vinylidene fluoride) (PVDF) in 1969, the fundamental physics of polymer piezoelectricity has remained elusive. Based on the classic principle of piezoelectricity, polymer piezoelectricity should originate from the polar crystalline phase. Surprisingly, the crystal contribution to the piezoelectric strain coefficient $d_{31}$ is determined to be less than 10%, primarily owing to the difficulty in changing the molecular bond lengths and bond angles. Instead, >85% contribution is from Poisson's ratio, which is closely related to the oriented amorphous fraction (OAF) in uniaxially stretched films of semicrystalline ferroelectric (FE) polymers. In this perspective, the semicrystalline structure–piezoelectric property relationship is revealed using PVDF-based FE polymers as a model system. In melt-processed FE polymers, the OAF is often present and links the crystalline lamellae to the isotropic amorphous fraction. Molecular dynamics simulations demonstrate that the electrostrictive conformation transformation of the OAF chains induces a polarization change upon the application of either a stress (the direct piezoelectric effect) or an electric field (the converse piezoelectric effect). Meanwhile, relaxor-like secondary crystals in OAF ($SC_{OAF}$), which are favored to grow in the extended-chain crystal (ECC) structure, can further enhance the piezoelectricity. However, the ECC structure is difficult to achieve in PVDF homopolymers without high-pressure crystallization. We have discovered that high-power ultrasonication can effectively induce $SC_{OAF}$ in PVDF homopolymers to improve its piezoelectric performance. Finally, we envision that the electrostrictive OAF mechanism should also be applicable for other FE polymers such as odd-numbered nylons and piezoelectric biopolymers.

# 1 | ELECTROSTRICTIVE ORIGIN OF PIEZOELECTRICITY IN FERROELECTRIC POLYMERS

Dielectric materials (or dielectrics) with a band gap greater than 3 eV are electric insulators with extremely low conductivity. They can be polarized by an external electric field to exhibit macroscopic polarization ($P$) with a high dielectric breakdown strength.[1,2] As such, they have found broad applications in electrical insulation (e.g., cables), energy storage (e.g., capacitors), and various electromechanical (e.g., piezoelectric/electrostrictive) and electrothermal coupling (e.g., pyroelectric/electrocaloric) devices.[3–10] The vast majority of dielectrics are linear dielectrics, whose dielectric constant (or relative permittivity, $\varepsilon_r$) can be defined via polarization: $P = (\varepsilon_r - 1)\varepsilon_0 E$, where $\varepsilon_0$ is the vacuum permittivity and $E$ is the applied electric field. In the theory of electrostatics, $P$ is also related to molecular structures via the equation: $P = M/V = N\mu = N\alpha E_L$, where $M$ is the macroscopic dipole moment, $V$ is the volume, $N$ is the dipole density, $\mu$ is the dipole moment of a dipole, $\alpha$ is the polarizability, and $E_L$ is the local electric field acting on the dipole, which is different from the applied electric field $E$. With these two equations, the macroscopic dielectric properties are closely related to their molecular structures,[1,2] and this is called the structure–dielectric property relationship. Dielectric materials include gases, liquids, and solids. Among the solids, they can be further divided into inorganics (e.g., ceramics), small molecules, and polymers, which can be either crystalline, semicrystalline, or amorphous.

Among dielectric materials, there is a subsidiary called piezoelectrics (Figure 1), which are usually crystalline in nature.[2,4,11] Among the total of 32 crystal classes based on point group symmetry, 10 classes are centrosymmetric (i.e., nonpolar) and 21 classes are non-centrosymmetric. Among these 21 non-centrosymmetric classes, 20 classes can exhibit direct piezoelectricity, and the remaining one has a cubic structure (i.e., nonpolar). Among the 20 piezoelectric classes, 10 classes are nonpolar but can exhibit direct piezoelectricity. Namely, their nonpolar crystal structure can be considered to transform into a "polar" one by applying an external stress, leading to direct piezoelectricity. A famous example of these direct piezoelectrics is quartz, which has been widely used in various sensors and transducers.[11–13] The other 10 classes are polar and thus exhibit spontaneous polarization ($P_s$) in the absence of any external electric field. These 10 classes also exhibit pyroelectricity, in which an electric polarization can be generated by changing the temperature. If the $P_s$ can be switched by reversing the direction of an external electric field, the material is ferroelectric (FE).

No known polymers belong to the 10 nonpolar direct piezoelectric classes because all piezoelectric polymers require $P_s$ to exhibit piezoelectricity. In their crystalline or oriented structures, a uniaxial symmetry is often present. In rectangular coordinates, the piezoelectric strain coefficient $d_{ij}$ (a tensor with $i = 1–3$ and $j = 1–6$) is defined as[2–4,11,14]:

$$d_{ij} = \left(\frac{\partial D_i}{\partial \sigma_j}\right)_E = \left(\frac{\partial \varepsilon_j}{\partial E_i}\right)_\sigma \quad (1)$$

where $D_i$ is the electric displacement, $\sigma_j$ is the stress, $\varepsilon_j$ is the strain, and $E_i$ is the electric field. The subscripts $E$ and $\sigma$ mean that the measurements are performed under constant $E$ and $\sigma$, respectively. The first equation defines the direct piezoelectricity and the second defines the converse piezoelectricity. Similarly, the piezoelectric stress coefficient, $e_{ij}$, is defined as[2–4,11,14]:

$$e_{ij} = \left(\frac{\partial D_i}{\partial \varepsilon_j}\right)_E = \left(\frac{\partial \sigma_j}{\partial E_i}\right)_\sigma \quad (2)$$

The **d** and **e** coefficients are related to each other via the elastic compliance, s: **d** = **es**, or the tensile modulus, **c**: **e** = **dc**. Here, **d**, **e**, **s**, and **c** are all tensors.

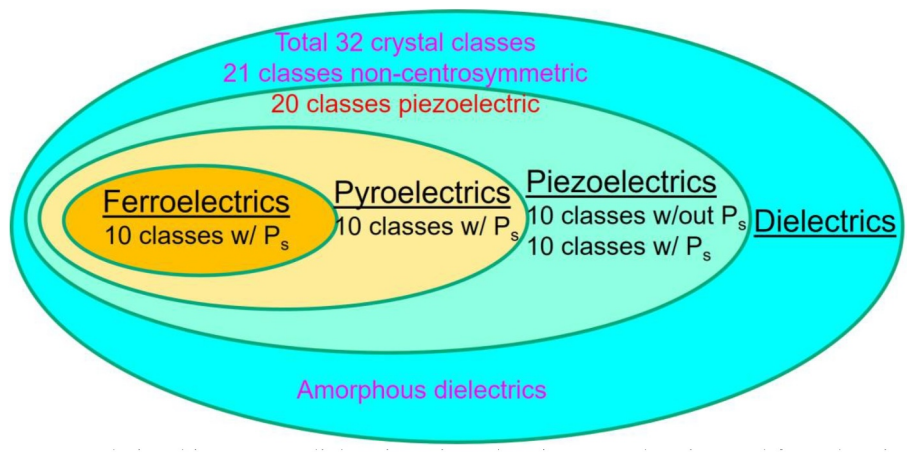

**FIGURE 1** Relationships among dielectrics, piezoelectrics, pyroelectrics, and ferroelectrics.

All piezoelectric polymers can be divided into two categories: biopolymers (non-ferroelectric) and poled FE polymers (with a permanent remanent polarization, $P_{r0}$). For biopolymers with either $\alpha$ helices or $\beta$ sheets, the 1 (the polarization) axis is perpendicular to the film plane and the 3 axis is the orientation axis. Shear piezoelectricity is universal for all biopolymers, which is along the 3* axis (i.e., 45° to the 3 axis; see Figure 2a).[15,16] For poled FE polymers [for example, poly(vinylidene fluoride) (PVDF) or odd-numbered nylons], the 3 ($P$ and $P_{r0}$) axis is perpendicular to the film plane and the 1 axis is the orientation axis (Figure 2b). Four different symmetries have been found for all piezoelectric polymers, relating to four $d_{ij}$ matrices.[15,16] For the $D_\infty$ symmetry in most natural (e.g., polypeptides) and synthetic [for example, poly(L-lactide) (PLLA) and polyhydroxyalkanoate (PHA)] biopolymers, the dipoles have an antiparallel orientation in the crystalline structure; therefore, no net dipole moment is present. In this case, only shear piezoelectricity exists with $d_{25} = -d_{14}$, and no pyroelectricity is possible. In certain structures of living systems (e.g., hair, bone, and tendon), the $C_\infty$ symmetry is observed. Moreover, the dipole moments orient uniformly in one direction. As such, both tensile and shear piezoelectricity are observed, together with a weak pyroelectric effect. The $C_{\infty v}$ symmetry is found for poled amorphous polar polymers[17,18] or polymer/FE ceramic particle composites[19–21] with $d_{24} = d_{15}$ (see Figure 2b). Finally, when uniaxially stretched FE polymers are poled, the $C_{2v}$ symmetry is obtained. Typical examples are PVDF (piezoelectricity was first discovered by Kawai in 1969,[22] and ferroelectricity was proved later by Davis in 1970[23]) and odd-numbered nylons (e.g., nylon-11, piezoelectricity was first discovered by Kawai in 1970,[24] and ferroelectricity was proved by Scheinbeim in 1984[25]).

For polar piezoelectrics with $P_s$, the origin of piezoelectricity is electrostriction,[3,14,26,27] which is ubiquitous for all dielectric materials. As shown in Figure 3a, dipoles (either induced or permanent) are random when there is no external electric field. After applying an electric field, they become oriented along the field direction, leading to electrostatic dipole–dipole interactions. These electrostatic interactions will change the shape of the dielectric mechanically, and this is called electrostriction. However, this effect is negligible for common dielectrics because the dipole–dipole interactions are too weak. Only for relaxor ferroelectrics (RFE) with high permittivity, is the electrostrictive strain ($S_{ij}$) large (up to ~0.5% for RFE ceramics[27] and up to ~7% for RFE polymers[3]). In theory, $S_{ij} = Q_{ijkl}P_kP_l$, where the electrostriction coefficient $Q_{ijkl}$ is a fourth-ranked tensor, relating to the second-order strain tensor $S_{ij}$ and the first-order polarization ($P_k$ and $P_l$). For isotropic materials, $P_k = P_l$ and thus $S_{ij} = Q_{ij}P^2$. A parabolic curve is observed for $S_{ij}$ as a function of $P$ in Figure 3b. When a bias polarization ($P_{bias}$, e.g., $P_{r0}$ in a poled FE sample) is present, another alternating polarization, $P_{ac} = (\varepsilon_r - 1)\varepsilon_0 E_0 e^{i\omega t}$, is applied.[26,27] Here, $E_0$ is the amplitude of the alternating electric field, $\omega$ is angular frequency, and t is time. The electrostrictive strain then becomes $S_{ij} = 2Q_{ij}P_{bias}P_{ac}$. By the definition of converse piezoelectric coefficient, we have:[3,27]

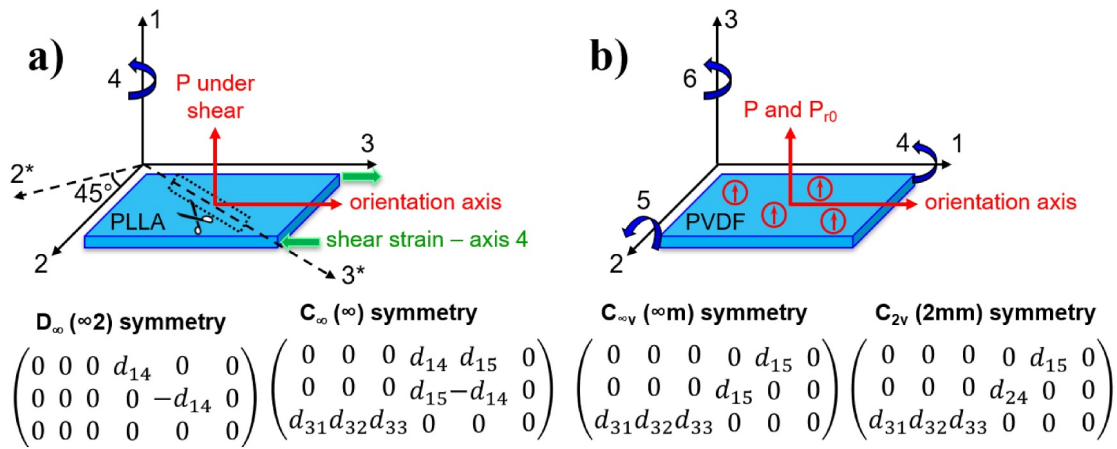

**FIGURE 2** Different piezoelectric tensors for (a) biopolymers (e.g., PLLA) with $D_\infty$ or $C_\infty$ symmetry and (b) poled ferroelectric polymers (e.g., PVDF) with $C_{\infty v}$ or $C_{2v}$ symmetry. In (a), shear piezoelectricity is induced by a stress along the 3* axis for biopolymers. In (b), the red circles with arrows represent either PVDF crystallites or ceramic particles in the PVDF matrix.

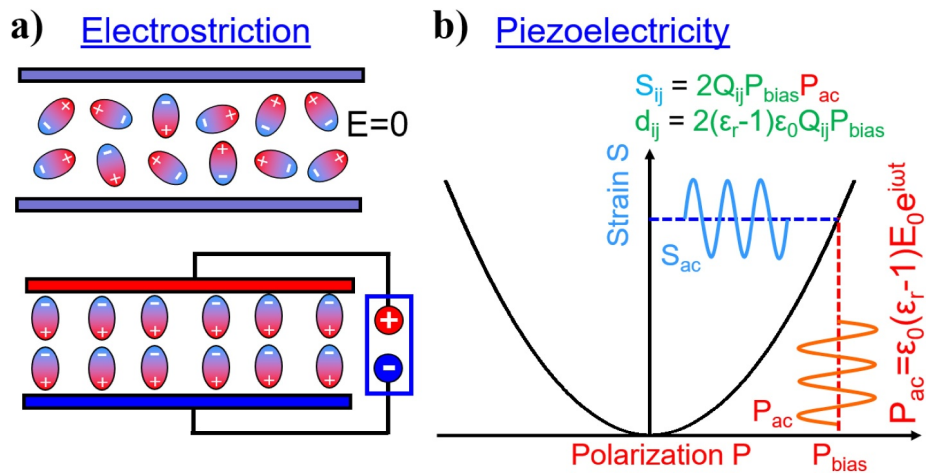

**FIGURE 3** Schematic representations of (a) electrostriction and (b) piezoelectricity.

$$d_{ij} = \partial S_{ij}/\partial E = 2Q_{ij}(\varepsilon_r - 1)\varepsilon_0 P_{bias} \quad (3)$$

Equation (3) signifies that piezoelectricity originates from the electrostriction of the dielectric material.

## 2 | MOTIVATION AND SCOPE OF THIS PERSPECTIVE ARTICLE

Compared to piezoelectric ceramic materials, whose $d_{ij}$ is ~200 pC/N for hard piezoelectrics and ~2000 pC/N for soft piezoelectrics,[28,29] piezoelectric polymers usually exhibit much lower $d_{ij}$ of less than 30 pC/N.[3,8,30,31] This has hampered the broad usage of piezoelectric polymers in real-world applications, for example, transducers in medical ultrasonic imaging and therapy.[32–34] It is highly desirable to enhance the piezoelectric performance of piezoelectric polymers to the levels of piezoelectric ceramics. However, piezoelectric polymers, which are mostly semicrystalline in nature, are very different from piezoelectric ceramics, where the morphotropic phase boundary (MPB) mechanism plays an important role.[35–40] Despite over 5 decades of extensive research, the fundamental mechanism of polymer piezoelectricity is still under debate: Whether the piezoelectricity is from the crystals, the amorphous phase, or the crystalline-amorphous interfaces? Without a comprehensive understanding of the complex semicrystalline structures in piezoelectric polymers, it is difficult to further enhance their piezoelectric performance. In this perspective, we will focus on poled FE polymers [that is, PVDF and P(VDF-co-trifluoroethylene) P(VDF-TrFE)][41] as an example to unravel the piezoelectric mechanism in solid films and further enhance their piezoelectric performance.

Piezoelectric biopolymers, including polysaccharides (cellulose, chitin, and amylose), proteins (collagen, keratin, and fibrin), DNA, and bio-polyesters (PLLA and PHAs)] are non-ferroelectric in nature. That is, the dipoles and polarization along the polymer chains in crystals cannot switch directions by applying an external electric field. However, we consider that the piezoelectric mechanism should be similar to that of FE piezoelectric polymers. Basically, electrostriction should be the origin of piezoelectricity. The current challenge is that the piezoelectric performance of biopolymers ($d_{14}$ = 0.1–10 pC/N) is significantly lower than that of FE piezoelectric polymers, such as PVDF and its random copolymers ($|d_{3j}|$ = 10–80 pC/N).[8–10] Readers can refer to other review articles on piezoelectric biopolymer solid films and their potential applications.[15,16,42–49] It is highly desirable to significantly enhance the piezoelectric performance for biopolymers.

In addition to solid films, fiber mats and foams of poled FE polymers and biopolymers can also exhibit piezoelectricity. However, if no electric poling is applied to FE polymers, either during or after the fiber-spinning process, no piezoelectricity should be observed. If an electric voltage is still observed for non-poled FE polymer fiber mats and foams, it is likely that the electric voltage is generated by triboelectricity (note that even rubbing the same material can also generate triboelectricity).[50] Electrospinning and electrospraying are powerful methods to generate macroscopic $P_{r0}$ (also called self-polarization) for FE polymer fiber mats,[51,52] and no post-poling is needed. By compressing the porous fiber mats and thus changing the volume and macroscopic polarization via $\Delta P = M(1/V_2 - 1/V_1)$, where $V_1$ and $V_2$ are volumes before and after compression, piezoelectricity is generated. Therefore, the working mechanism for electrospun FE polymer fiber mats is simply the dimensional (or composite) effect, which will be discussed later. For piezoelectric biopolymers, a similar situation is found. Conventional biopolymer fiber mats (e.g., obtained from melt and solution fiber-spinning processes) does not exhibit any piezoelectricity without electrospinning because the dipole moments in the crystalline fibers (assuming that the α-helices are along the fiber direction) will cancel each other (i.e., the $D_\infty$ symmetry). As a result, all tensile piezoelectric coefficients are zero and only the shear piezoelectric coefficient $d_{14}$ exists. However, given the fiber geometry with a small diameter, shear stresses are difficult to apply at a 45° angle to the fiber axis (see Figure 2a). Therefore, the piezoelectricity in non-electrospun biopolymer fiber mats should be zero or very weak. However, electrospinning can effectively break the $D_\infty$ symmetry and realize the $C_\infty$ symmetry for biopolymer fiber mats. As a result, tensile piezoelectric coefficients are then nonvanishing, and piezoelectricity is often observed when a stress is applied along the fiber direction. Examples include electrospun PLLA,[53,54] PHAs,[55,56] and poly(γ-benzyl L-glutamate) (PBLG).[57] These piezoelectric fiber mats can find potential applications in scaffolds for bone regeneration and tissue engineering, air filtration such as N95 masks, and parasitic mechanical energy harvesting. For electrospun piezoelectric polymer fiber mats and their potential applications, readers can refer to recent review articles.[44,46,48,49,58–64]

## 3 | COMPLEX STRUCTURES OF SEMICRYSTALLINE POLYMERS—EXISTENCE OF THE ORIENTED AMORPHOUS FRACTION

In this section, we will review the complex structures of semicrystalline polymers and their relationships to piezoelectricity. In 1928, Meyer and Mark proposed the fringed-micelle model from the fiber wide-angle X-ray diffraction (WAXD) study of crystalline cellulose.[65] In this model, the crystal stem length is significantly shorter than the entire chain length, and thus the amorphous tie chains should connect between neighboring crystallites. In 1938, Storks studied thin films of gutta-percha (i.e., trans-1,4-polyisoprene) using electron diffraction.[66] He proposed that the polymer chains had to fold back and forth to fit the thin film thickness. In 1957, Keller unambiguously proved chain-folding for thin polyethylene single crystals using transmission electron microscopy and electron diffraction.[67] Similar results were obtained independently by Fisher and Till in the same year.[68,69] By etching polyethylene single crystals using fuming nitric acid followed by a size-exclusion chromatography study, it was confirmed that the chain-folding should adopt adjacent re-entry.[70] Since then, it is widely accepted that chain-folding is the fundamental mechanism for polymer crystallization for kinetic reasons [for example, the barrier to extended-chain crystals (ECCs) is too high].[71–73]

Chain-folding is observed in polymer single crystals, which are grown from dilute solutions. However, the crystallization mechanism may be different when a polymer is crystallized from the melt, especially under processing conditions, such as fiber spinning, film extrusion, injection molding, and stretch blow molding. As shown in Figure 4a, folded-chain crystals (FCCs) are often obtained for dilute solution-grown polymer single crystals. For crystals produced from concentrated solutions or melts, not all chains undergo chain-folding due to chain entanglement and high melt viscosity during crystallization. There should be a significant portion of polymer chains protruding out of the crystalline lamellae, forming the oriented amorphous fraction (OAF) caused by chain crowding at the crystalline–amorphous interface (Figure 4b). As the chains move away from the crystalline–amorphous interface, their orientation is gradually lost, and they enter the isotropic amorphous fraction (IAF). For highly stretched polymer fibers and films, the situation can be further complicated by the formation of microfibrillar crystals (Figure 4c). Within the fibrillar crystals, there could be taut-tie molecules or non-crystalline defects (which can be considered as a special type of OAF) between lamellar crystallites. Between different fibrillar crystals, there are OAF and IAF, as well as long-chain tie molecules. Because the OAF has one end tethered on the solid crystals in Figure 4b, the OAF chains can become rigid; therefore, a significant portion of OAF becomes the rigid amorphous fraction (RAF). The IAF is mobile and can be considered as the mobile amorphous fraction (MAF). Differential scanning calorimetry (DSC), especially temperature-modulated DSC (TMDSC), has been widely used to determine the content of RAF in various polymers.[74,75] However, the OAF content could not be determined by DSC, and we need to resort to fiber WAXD analysis, as detailed by Wunderlich and coworkers.[76–78] The relationship between crystals, OAF/RAF, and IAF/MAF is shown in Figure 4d. In general, DSC can detect both large and poor crystals, whereas WAXD cannot accurately detect poor and mesomorphic crystals. Therefore, the crystallinity ($x_c$) determined by DSC is usually higher than that determined by WAXD. Because fiber WAXD cannot accurately detect poor and mesomorphic crystals, the IAF content ($x_{IAF}$) is often overestimated as compared to the MAF content ($x_{MAF}$) determined by DSC. As a result, the contents of OAF ($x_{OAF}$) and RAF ($x_{RAF}$) may or may not be identical.

## 4 | DIFFERENT CONTRIBUTIONS TO POLYMER PIEZOELECTRICITY

Given different components in semicrystalline polymers, we ask a question: What are their contributions to polymer piezoelectricity? Based on the equations: $d_{3j} = \partial P_3/\partial \sigma_j$, $e_{3j} = \partial P_3/\partial \varepsilon_j$, and $P_3 = M_3/V$, $d_{3j}$ and $e_{3j}$ can be rewritten as the following (see Figure 5a):[79,80]

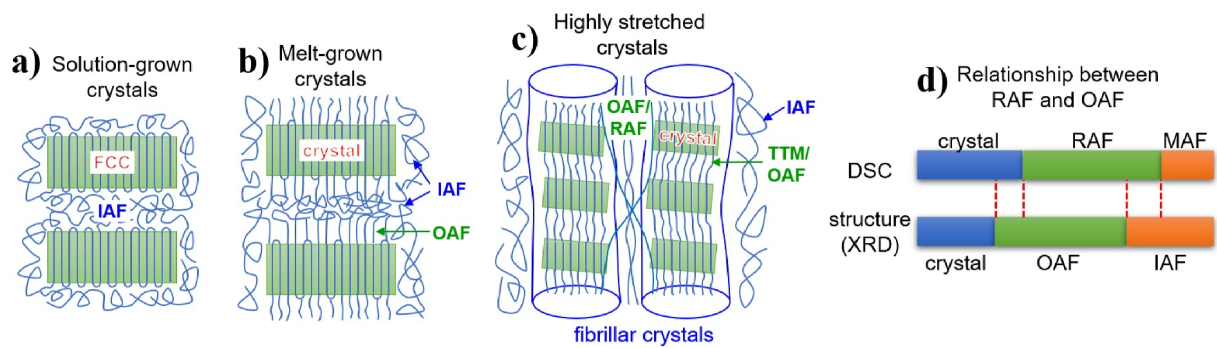

**FIGURE 4** Complex structures of semicrystalline polymers: (a) solution-grown folded-chain crystals (FCC), (b) melt-grown lamellar crystals with oriented and isotropic amorphous fractions (OAF and IAF), (c) highly stretched fibrillar crystals with taut-tie (TTM, intra-fibrillar crystal) and long-chain (inter-fibrillar crystal) tie molecules, OAF, and IAF, and (d) shows the relationship between crystals, rigid amorphous fraction (RAF)/OAF, and mobile amorphous fraction (MAF)/IAF determined by DSC and XRD, respectively.

$$Y_{3j} = P_3\left(\frac{\partial \ln M_3}{\partial X_j} - \frac{\partial \varepsilon_3}{\partial \varepsilon_j}s_j\right), Y = d \text{ or } e \text{ and } X = \sigma \text{ or } \varepsilon \quad (4)$$

where $P_3$ and $M_3$ are the spontaneous polarization and the macroscopic dipole moment along the 3 direction, respectively, and $s_j$ is the compliance along the $j$ ($j = 1, 2, 3$) direction. $P_3$ comes from the poled FE crystals. Without any FE crystals, $P_3$ does not exist, and no piezoelectricity can be detected. The second terms in the parentheses tell us that one contribution is from the compliance of the material. For hard piezoelectric ceramics, the compliance is negligible. For soft polymers, the compliance, which originates from the amorphous phase above the glass transition temperature ($T_g$), is relatively high; the higher the compliance, the higher the $d_{3j}$ and $e_{3j}$. The first term in the parentheses is the major contribution to piezoelectricity for both ceramic and polymeric piezoelectrics, namely the macroscopic dipole moment $M_3$, which increases in response to the applied stress $\sigma_j$ (for $d_{3j}$) or strain $\varepsilon_j$ (for $e_{3j}$). For ceramic piezoelectrics such as lead zirconate titanate, this term was first thought to originate from the easy crystal–crystal transformation between FE rhombohedral and the tetragonal phases at the MPB.[35–40] Later, it was found that domain walls also play an important role.[81,82]

For polymeric piezoelectrics, the situation must be different from that of ceramic piezoelectrics. This then raises the question: Which component contributes significantly to the first term in the parentheses of Equation (4): the FE crystals or the crystalline–amorphous interfaces? Let us first assume it is the crystalline phase. For rigid crystals such as PVDF, breathing motions (i.e., changes of $CF_2$ and C-C-C bond angles) of the polymer chains are proposed to explain the positive $e_{31}$ and negative $e_{33}$.[80,83] However, it is known that bond angles of polymers are extremely difficult to change with external stresses/electric fields, and in situ WAXD has already shown that no observable changes in unit cell dimensions could be observed during the application of an external stress[84] or an external electric field to the piezoelectric PVDF.[85] Therefore, this molecular model cannot explain polymer piezoelectricity.

Instead, polymer piezoelectricity should be explained using chain conformation changes. As shown in Figure 6a,[86] without any electric field (i.e., $E = 0$), the polymer crystal adopts a twisted conformation, which is similar to that in the paraelectric (PE) crystals of P(VDF-TrFE) random copolymers.[89,90] Upon application of an external electric field ($E > 0$) of sufficient strength, the chain conformation in the crystals transforms into the all-trans conformation with dipoles aligning in the field direction. As a result, $S_1$ along the chains will increase and $S_3$ in the thickness direction will decrease. Here, let us use the P(VDF-TrFE) 55/45 mol.% sample as an example to estimate the changes in both $S_1$ and $S_3$. By considering the unit cell dimensions of the PE and FE crystals,[89,90] the $S_1$ for the PE (001) to FE(001) transformation is found to be 10.9%. If we take into account that the $b$-axis of the FE phase (i.e., the dipole direction) is oriented in the normal direction of the poled crystal, the $S_3$ for the PE(010) to FE(010) transformation is −3.3%. If we do not consider dipole orientation in the $b$-axis and only consider the PE(110) to FE(110) transformation then $S_3$ is −6.7%.

Using combined density functional theory (DFT) and molecular dynamics (MD) models, Strachan and Goddard set up a crystalline slab for PVDF, as shown in Figure 6b.[87] Due to the electric field-induced transformation from the twisted to the all-trans conformation, a maximum electrostriction of around 5% was obtained with $Q_{33} = 2.5$–$3.0$ m$^4$/C$^2$ in the GHz range (Figure 6c). Assuming $\varepsilon_r = 12$ and $P_{bias} = P_{r0} \sim 120$ mC/m$^2$,[91] the $-d_{33}$ can be calculated to be 58–70 pC/N from Equation (3). By studying a series of P(VDF-co-trifluoroethylene) [P(VDF-TrFE)] random copolymers with the VDF content around 50 mol.%, Liu et al. proposed that piezoelectricity originates from the crystalline phase with an MPB-like mechanism.[8,88,92–94] For P(VDF-TrFE) random copolymers, the MPB was found to be around VDF% ~50 mol.%. Below 50 mol.% VDF, the polymer chains tended to adopt a helical (or twisted) conformation in the crystals. Above 50 mol.% VDF, the polymer chains tended to adopt an all-trans conformation in the crystals. By measuring the $S_3$-$E$ loops for annealed and poled samples (unstretched), $d_{33}$ values were obtained from the average

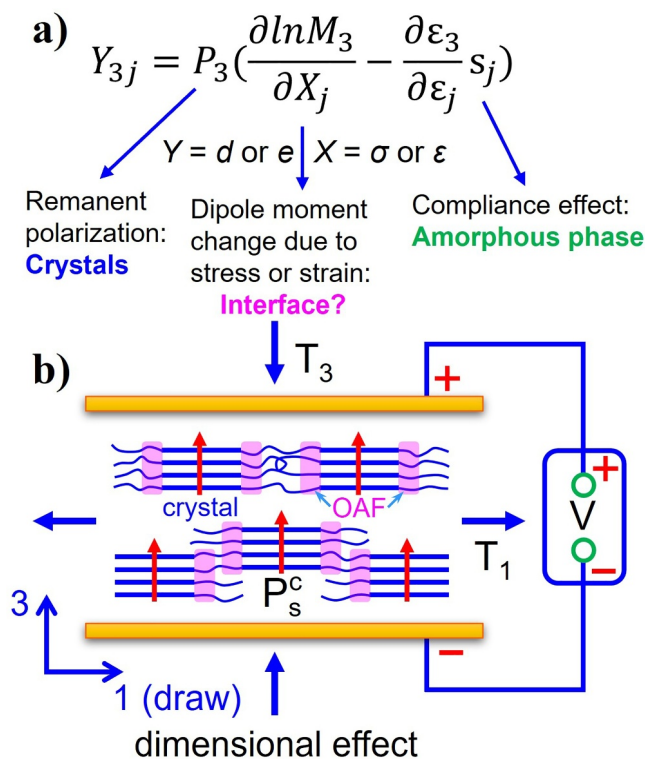

**FIGURE 5** (a) Different contributions to piezoelectric constants $d_{3j}$ or $e_{3j}$. (b) Schematic representation of the dimensional effect for polymer piezoelectricity using a composite model, that is, oriented crystallites (chains along the 1 direction) in the amorphous polymer matrix.

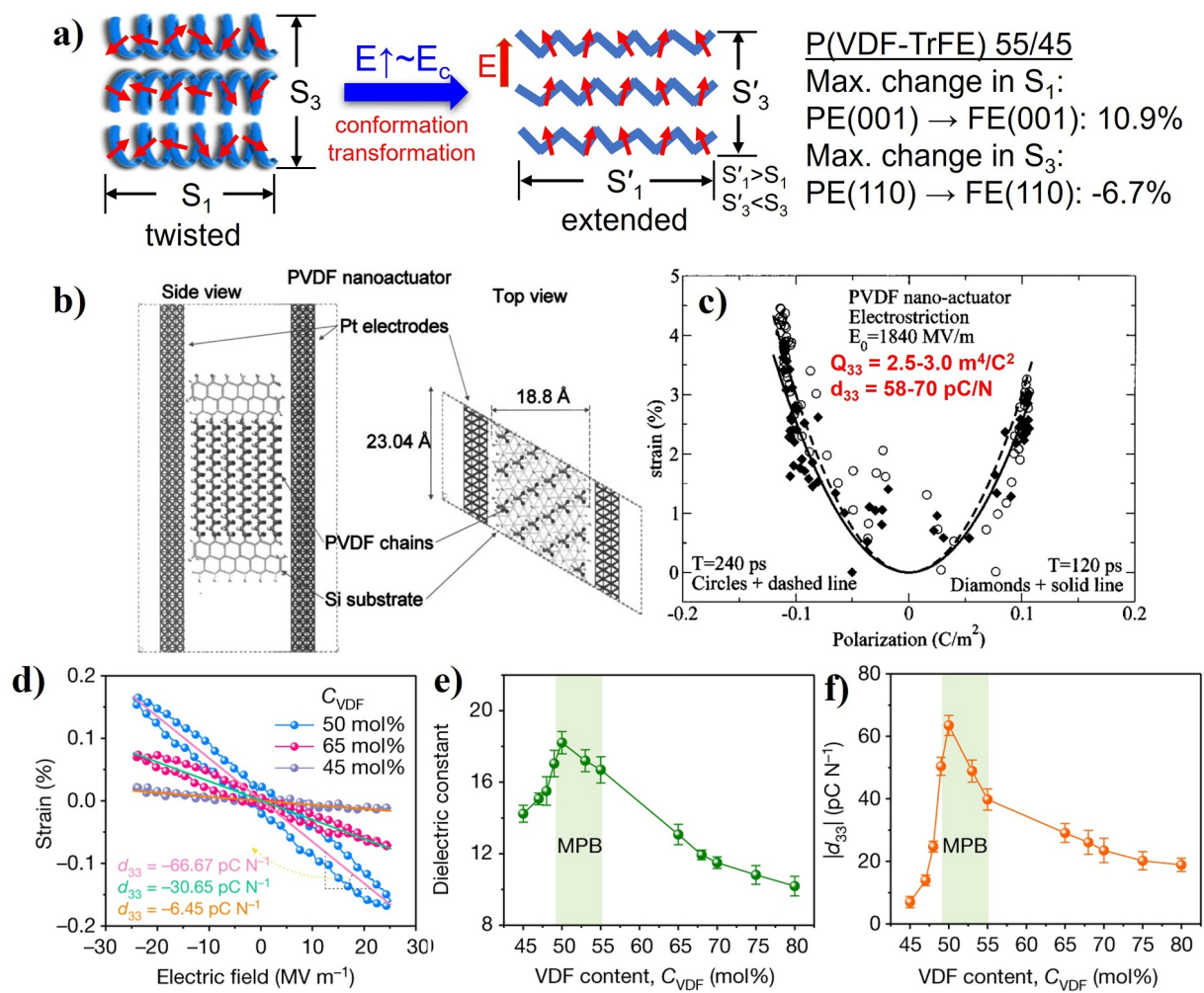

**FIGURE 6** (a) Schematic representation of the conformation transformation (i.e., from the twisted to the all-trans conformation) in the crystalline phase when an external electric field is applied, resulting in a positive strain $S_1$ and a negative strain $S_3$. Reproduced with permission.[86] Copyright 2020, American Chemical Society. (b) Simulation setup for the nano-actuation of a PVDF slab of the β crystal and (c) simulated strain as a function of polarization at 4.2–8.3 GHz. Reproduced with permission.[87] Copyright 2005, AIP Publishing. (d) Strain $S_3$ as a function of poling field and (e) dielectric constant, and (f) $|d_{33}|$ as a function of the VDF content. Reproduced with permission.[88] Copyright 2018, Springer Nature.

slopes (Figure 6d). Both dielectric constant (Figure 6e) and $|d_{33}|$ (Figure 6f) reached the maximum values around 50/50 mol.% composition. It was considered that the easy conformation transformation between the helical and all-trans conformations around the MPB enhanced both the dielectric constant and $|d_{33}|$.

Although the MPB mechanism can be used to explain the enhanced piezoelectric behavior of P(VDF-TrFE) around 50/50 mol.%, it cannot explain the piezoelectricity for PVDF homopolymers because there is no such a crystal–crystal transformation for PVDF homopolymers. For example, careful WAXD studies showed negligible crystal contribution to $d_{31}$ but significant crystal contribution to $d_{33}$ for the neat PVDF homopolymer.[84] To detect whether there is any crystal–crystal transformation for P(VDF-TrFE) random copolymers, in situ two-dimensional (2D) WAXD is needed when either a stress or an electric field is applied to the piezoelectric polymer samples. Currently, research is underway and will be reported in the future.

To explain the experimental $d_{31}$ values of PVDF, the amorphous phase needs to be added as the matrix for poled FE crystals. Figure 5b shows a composite or dimensional model.[84,95–100] In this model, poled FE crystals are dispersed in an amorphous PVDF matrix. Wada and coworkers developed a theoretical treatment for this dimensional model.[95,97] Later, Tashiro and Tadokoro further detailed this theory and calculated macroscopic $d_{31}$ and $d_{33}$ values for PVDF and their limiting values.[84,98] Below, we will briefly review the theoretical treatment of the dimensional model and discuss its relationship to the electrostriction origin for polymer piezoelectricity.

As shown in Figure 5b, poled FE crystals (with chains along the 1 or stretching direction) are dispersed in the amorphous polymer matrix. The dielectric constants of the amorphous phase and the FE crystals are $\varepsilon_r^a$ and $\varepsilon_r^c$, respectively. The $P_s$ of the poled crystal is $P_s^c$; therefore, the $P_s$ of the composite ($P_s^M$) can be obtained as $P_s^M = 3\phi\, P_s^c \varepsilon_r^a/(2\varepsilon_r^a + \varepsilon_r^c)$.

Starting from the definition of the piezoelectric stress coefficient $e_{3j}$:[84,95,97,98]

$$e_{3j} = \frac{1}{A}\left(\frac{\partial Q_3}{\partial \varepsilon_j}\right)_{E=0}, \qquad (5)$$

where $A$ is the sample area and $Q_3$ is the surface charge, we can reach a final expression for $e_{3j}$:[84,95,97,98]

$$e_{3j} = \frac{3\phi\varepsilon_r^a}{2\varepsilon_r^a + \varepsilon_r^c}\left\{P_s^c\left[\frac{\varepsilon_r^c}{2\varepsilon_r^a + \varepsilon_r^c}\left(\frac{\kappa_{3j}^a}{\varepsilon_r^a} - \frac{\kappa_{3j}^c}{\varepsilon_r^c}\right) + v_{3j}\right.\right.$$
$$\left.\left. + \frac{1}{V_c}\frac{\partial V_c}{\partial \varepsilon_j}\right] + \frac{\partial P_s^c}{\partial \varepsilon_j}\right\} \qquad (6)$$

where $\kappa_{3j}^a$ and $\kappa_{3j}^c$ are dielectrostriction constants: $\kappa_{3j}^{a/c} = \partial \varepsilon_r^{a/c}/\partial \varepsilon_j$. We call it *dielectrostriction* in order to differentiate it from the true electrostriction term mentioned above (see Figure 3 and corresponding discussion). In general, $\kappa_{3j}^c$ is significantly smaller than $\kappa_{3j}^a$ and thus can be ignored. $v_{3j}$ is the macroscopic Poisson's ratio: $v_{3j} = -\partial \ln t/\partial \varepsilon_j = -\partial \varepsilon_3/\partial \varepsilon_j$, where $t$ is the film thickness. Note that $v_{33} = 1$. $V_c$ is the volume of the FE crystal, and $\phi$ is the volumetric crystallinity. Here, a new definition is used for $e_{3j}$ with Equation (6), rather than Equation (2) because the sample area changes when a strain is applied.[101] If the sample area does not change, Equations (2) and (6) give the same result for $e_{3j}$. From Equation (6), four contributions are identified for polymer piezoelectricity: (i) dielectrostriction, (ii) Poisson's ratio, (iii) volume change of the crystal, and (iv) $P_s^c$ change of the crystal (i.e., intrinsic piezoelectricity of the FE polymer crystals). Taking into account various the physical properties of uniaxially stretched PVDF, the $e_{3j}$ and $d_{3j}$ values at room temperature (RT) are listed in Table 1. As we can see, $v_{3j}$ ($v_{31}$ = 0.6–0.7[104,105] and $v_{33}$ = 1) is the major contributor (>85%) to polymer piezoelectricity. This is consistent with the conclusion of an earlier study by Sussner that the temperature-dependent Poisson's ratio should be used to explain the sudden drop of piezoelectric coefficients below the $T_g$ of PVDF.[104]

During the $e_{3j}$ and $d_{3j}$ calculation for uniaxially stretched PVDF, several factors are noticed. (i) The higher the $v_{3j}$, the higher $e_{3j}$ and $d_{3j}$. Based on a mechanical property study of low-density polyethylene, $v_{31}$ can reach as high as 0.8 at RT.[106] (ii) The volumetric crystallinity $\phi$ plays an important role. When $\phi = 0$, $e_{3j} = 0$. When $\phi = 1$, $e_{3j}$ and $d_{3j}$ become the intrinsic piezoelectric property of the FE PVDF crystal, which is negligibly small. Therefore, the highest $e_{3j}$ and $d_{3j}$ should happen around $\phi \sim 0.5$. (iii) $e_{3j}$ is almost independent of the Young's modulus $c_j$, whereas $d_{3j}$ is inversely proportional to $c_j$ because $d_{3j} = e_{3j}/c_j$. Therefore, to get a high $d_{3j}$, $c_j$ should not be high. Using the following parameters: $v_{31} = 0.8$, $\phi = 0.5$, $c_1 = 2.2$ GPa, $c_2 = c_3 = 1.6$ GPa, and perfect crystal orientation (i.e., b-axis along the 3 direction), limiting values of $e_{3j}$ and $d_{3j}$ are calculated and listed in Table 1.

It is known that the highest Poisson's ratio for isotropic amorphous polymers is 0.5 for natural rubbers.[107] The values of $v_{31}$ higher than 0.5 for uniaxially stretched PVDF[104,105] and LDPE[106] are interesting observations. If we use $v_{31} = 0.3$ for unstretched PVDF, $d_{31}$ should only be ~10 pC/N. The question is: Why is $v_{31}$ higher than 0.5 for uniaxially stretched semicrystalline polymers? Tasaka and Miyata observed that the birefringence of rolled PVDF films continuously increased with increasing the roll ratio, whereas the crystal orientation factor stopped increasing beyond a roll ratio of 2.[105] Meanwhile, both the

**TABLE 1** Calculated room temperature (RT) and limiting $e_{3j}$ and $d_{3j}$ values for uniaxially stretched and poled PVDF with breakdown of different contributions.

| $e_{3j}$ (mC/m$^2$) $d_{3j}$ (pC/N) | Observed | Calculated | % Contributions | | | |
|---|---|---|---|---|---|---|
| | | | $\kappa_{3j}^a$ | $v_{3j}$ | $\partial \ln(V_c)/\partial \varepsilon_j$ | $\partial P_s^c/\partial \varepsilon_j$ |
| $e_{31}$ (RT) | ~40[78,96] | 28.7 | 11.9 | 87.8 | 0.8 | −0.5 |
| $-e_{32}$ (RT) | ~5[78,96] | 7.2 | −4.8 | 100.1 | 27.9 | −23.2 |
| $-e_{33}$ (RT) | 40–105[78,96] | 40.7 | −1.7 | 88.5 | −12.3 | 25.5 |
| $d_{31}$ (RT) | 20–75[96,102] | 25.3 | 5.4 | 87.9 | −6.6 | 13.4 |
| $-d_{32}$ (RT) | ~2[96] | 7.0 | −8.8 | 96.2 | 8.7 | 3.8 |
| $-d_{33}$ (RT) | 20–65[96,103] | 35.4 | 1.7 | 88.6 | −7.9 | 17.6 |
| $e_{31}$ (limit) | – | 161.1 | 10.6 | 89.3 | 0.6 | −0.4 |
| $-e_{32}$ (limit) | – | 17.8 | −10.1 | 101.0 | 55.6 | −46.5 |
| $-e_{33}$ (limit) | – | 202.5 | −1.8 | 88.8 | −12.4 | 25.4 |
| $d_{31}$ (limit) | – | 144.9 | 4.4 | 86.7 | −6.1 | 14.8 |
| $-d_{32}$ (limit) | – | 16.7 | −12.7 | 96.8 | 27.5 | −11.6 |
| $-d_{33}$ (limit) | – | 186.3 | 2.0 | 88.9 | −8.0 | 16.9 |

*Note*: Reproduced with permission.[96] Copyright 1983 American Chemical Society.

dielectrostriction constant $\kappa_{31}$ and $\nu_{31}$ increased with increasing birefringence of the uniaxially stretched PVDF films. They suggested that the orientation of the amorphous phase must have had an important effect on $\nu_{31}$ and thus on the piezoelectricity of PVDF. In addition, it has been observed that $d_{3j}$ often has a stepwise increase and $e_{3j}$ has a stepwise decrease as the temperature passes through the $T_g$ of various polymers, including PVDF ($T_g \sim -45°C$), nylon-11 ($T_g \sim 45°C$), PLLA ($T_g \sim 55°C$), and PHB ($T_g \sim 5°C$).[15,26,104,108] It again indicates that "the oriented amorphous phase" must have played an important role in polymer piezoelectricity (note: the isotropic amorphous phase should not exhibit any piezoelectricity due to its nonpolar structure). However, it is never clearly pointed out where "the oriented amorphous phase" is in semicrystalline piezoelectric polymers.

## 5 | ROLE OF INTERFACIAL OAF ON PIEZOELECTRIC PROPERTIES OF POLYMERS

Harnischfeger and Jungnickel studied the dynamic and nonlinear piezoelectric properties of PVDF.[103] Dynamic relaxations and nonlinearity were observed for the piezoelectric PVDF samples. Three major conclusions were drawn from their study: (i) All piezoelectric relaxations could occur only in noncrystalline regions. (ii) Nonlinear piezoelectricity was possible only via a stress-induced change of molecular dipoles in noncrystalline regions, and (iii) the crystalline phase could not account for any nonlinearity. These conclusions contradict the conventional opinion that polymer piezoelectricity should come from the crystalline phase. To resolve this contradiction, they proposed a transition phase or interphase between the crystalline and amorphous phases. Both intrafibrillar and interfibrillar interphases could exist in semicrystalline PVDF. In contrast to the behavior of crystals, these interphases had higher dipole rotational mobility (which is similar to that of the amorphous phase), but still had a certain molecular orientation due to the tethering with the crystalline phase. It should be these interphases that enabled the relaxational and nonlinear piezoelectricity for semicrystalline PVDF. It is the first time that the crystalline–amorphous interface (i.e., the OAF) was clearly proposed to explain the piezoelectricity of PVDF. We consider that it is also the determining factor for the Poisson's ratio of uniaxially stretched semicrystalline polymers. Recently, the crystalline–amorphous coupling was further inferred from the piezoelectric strain coefficient $d_{33}$ obtained during in situ electric poling of a P(VDF-TrFE) 70/30 mol.% random copolymer.[109]

Based on the complex semicrystalline morphology in Section 2 and the above studies on crystalline–amorphous interfacial contribution,[103,109] we propose the OAF mechanisms for the direct and converse piezoelectricity of polymers. As shown in Figure 7, a significant portion of the PVDF chains are pulled out from the FE crystals by the uniaxial mechanical stretching, leading to OAF formation.[110,111] The OAF resembles a liquid crystalline polymer having a chain orientation, but the crystalline registry is largely lost. As a result, it could not give well-defined X-ray reflections. We have demonstrated that the VDF dipoles are highly mobile in the OAF at RT.[112] The reorientation of VDF dipoles in the OAF leads to the electrostriction effect, enhancing the piezoelectric performance. Namely, electrostriction is the origin of polymer piezoelectricity.[14,26,27] For example, when a stress is applied along the chain direction, the sample elongates with a strain $\delta$. As a result, more dipoles in the OAF will flip up to induce an additional polarization, $\Delta P$. This is the direct piezoelectric effect, leading to a positive $d_{31}$ and a negative $d_{33}$: $P_3 = d_{3j}\sigma_j$ (Figure 7a). Conversely, when a positive electric field is applied, the dipoles in the OAF will align in the up direction. The electrostatic repulsion among the positively aligned crystalline and OAF domains leads to an elongation in the horizontal direction. When a negative electric field is applied, the dipoles in the OAF will align in the downward direction. The electrostatic attraction among oppositely aligned crystalline and OAF domains leads to a shrinkage in the horizontal direction. This is the converse piezoelectric effect: $S_j = d_{3j}E_3$ (Figure 7b). As mentioned above, within the linear piezoelectric (i.e., linear dielectric and mechanical) regime, direct $d_{3j}$ (unit pC/N) and converse $d_{3j}$ (unit pm/V) are equal. Note that in SI units, pC/N = pm/V = A s$^3$ kg$^{-1}$ m$^{-1}$.

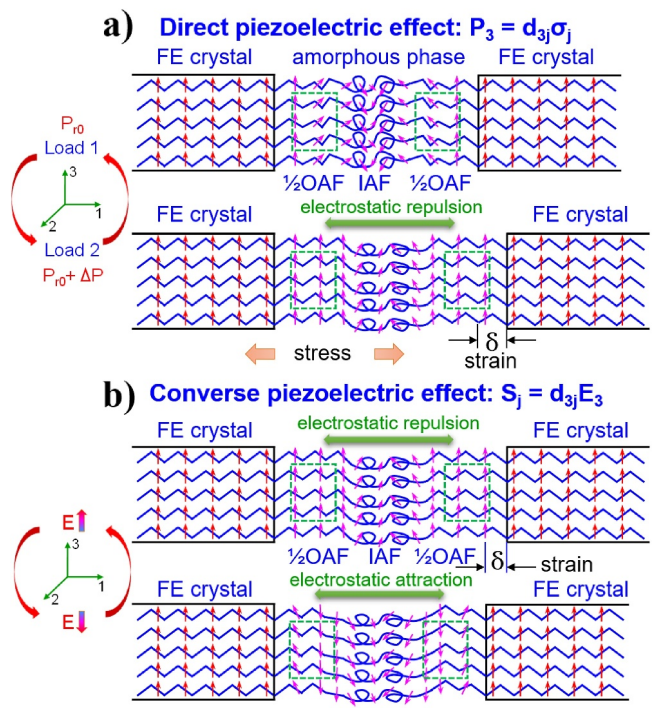

**FIGURE 7** Schematic representations of the (a) direct and (b) converse piezoelectric effects. The green dashed boxes represent the secondary crystals in the OAF. Reproduced with permission.[110] Copyright 2021, Elsevier, Inc.

This electrostriction model is not in conflict with the dimensional model in Section 3. When there is no OAF in the dimensional model, it is a pure two-phase composite model with only poled FE crystals in the isotropic amorphous matrix. As such, the Poisson's ratio $v_{31}$ can be as low as 0.3, and the $d_{31}$ of PVDF can only be 10 pC/N. When there is a significant amount of OAF in the dimensional model (see Figure 5b), the Poisson's ratio $v_{31}$ can increase to 0.7, and the $d_{31}$ is as high as 25 pC/N.[84] Consequently, the fundamental principle of the dimensional model is electrostriction (i.e., the electric field-induced conformation transformation in the OAF). This was also pointed out by Furukawa in the past.[113] However, the exact relationship between the OAF and the high $v_{31}$ for highly stretched semicrystalline polymers remains to be unraveled. Currently, research is underway.

To add detail to the schematic representations in Figure 7, we also carried out MD simulations to quantify the OAF contribution to both direct and converse piezoelectricity in PVDF (Figure 8).[110,111] Altogether 12 × 12 PVDF chains are oriented in the x-direction with each chain containing 30 repeat units. For both ends on the left and right walls, the VDF dipoles are fixed upward along the y-direction. These chains can be considered as ideal OAF between two poled PVDF crystals without any IAF in the middle. In this simulation, there is no need to implement the poled PVDF crystals. Instead, we only use the left and right walls with fixed dipoles to represent the surfaces of poled $\beta$ crystals. When we set the ab unit cell dimensions to those of the $\beta$ phase ($a = 0.858$ nm and $b = 0.491$ nm), the MD simulation takes an extremely long time to equilibrate due to high dipolar interactions among the chains. We therefore enlarge the ab unit cell dimensions to $a = 2.1$ nm and $b = 1.2$ nm to decrease the interactions and speed up the equilibration process for the simulation of the direct piezoelectric effect (Figure 8a–c).[111] As shown in Figure 8a, as the strain $S_1$ along the PVDF chains increases, more dipoles near the left and right walls become red with a positive dipole moment of 2.1 D. The average thickness of the upward dipoles in the OAF at the crystalline–amorphous interfaces ($t_{OAF,up}$) increases with increasing $S_1$ (Figure 8b). Meanwhile, we calculate the polarization values of the first 4 and 8 repeat units from both walls, and they increase with increasing $S_1$, exactly indicating the direct piezoelectric effect: $P_3 = d_{31}\sigma_1 = d_{31}c_1S_1$. This simulation result clearly identifies the origin of the first term in the parenthesis of Equation (4). Namely, the conformation transformation (or electrostriction) is the major contribution to polymer piezoelectricity.

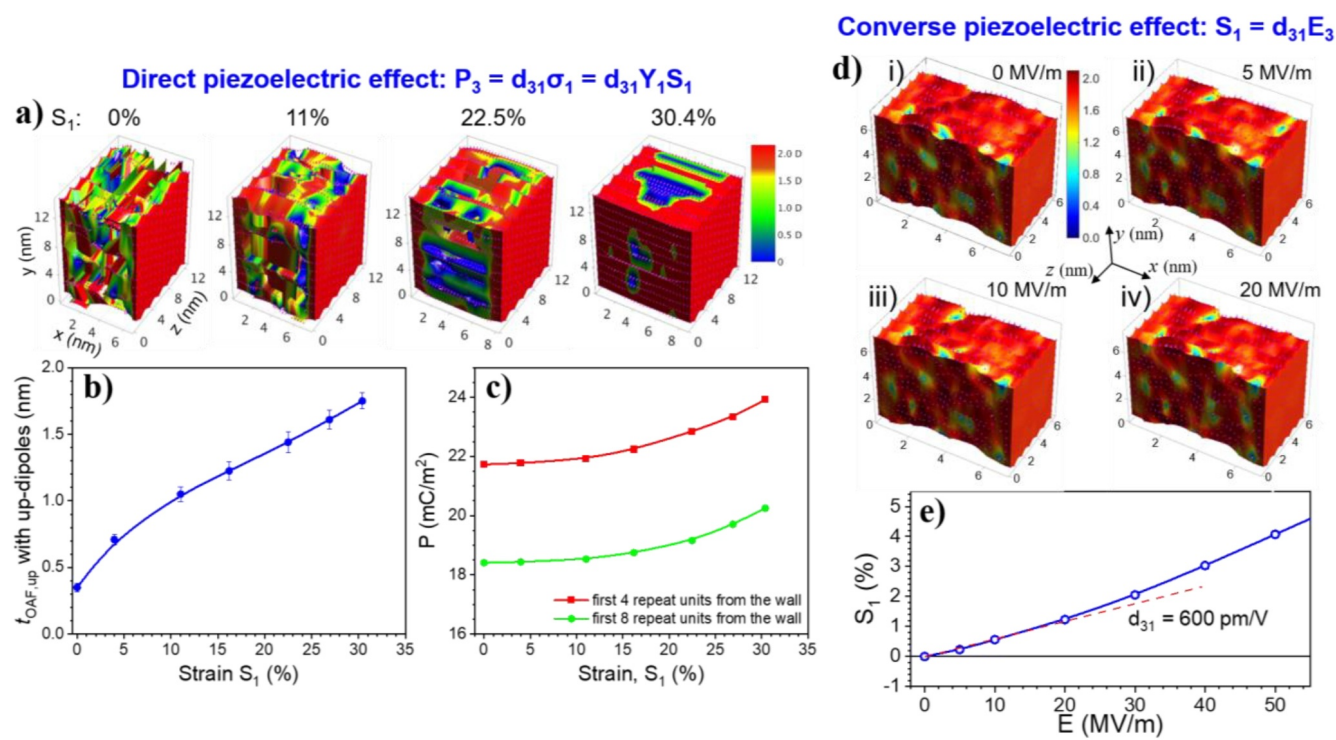

**FIGURE 8** (a) Simulation slabs for PVDF after equilibration at 300 K for 2 ns. The color scale represents PVDF units with positive dipole moments along the y-axis with values between zero (blue color) and 2.1 D (red color). The empty space in the pictures is composed of negative dipoles. There are 12 chains along the y-axis, 12 chains along the z-axis, and 30 repeat units along the x-axis. The slab thickness along x is 6.8 nm. Chain ends are attached to both slab walls with a rigid C-C bond having the dipole moment fixed along y. The attachment points are organized in the same manner as the $\beta$ crystal. The ab unit cell dimensions on the slab wall are $a = 2.1$ nm and $b = 1.2$ nm, respectively. The strain along x ($S_1$) is 0%, 11%, 22.5%, and 30.4%, respectively. (b) The average thickness of the upward dipoles in OAF ($t_{OAF,up}$) and (c) the polarization along y (P) for the first 4 and 8 repeat units from the wall as a function of $S_1$. Reproduced with permission.[111] Copyright 2021, Springer Nature. (d) Simulation slabs for PVDF under (i) 0, (ii) 5, (iii) 10, and (iv) 20 MV/m at 300 K. The slabs are the same as those in (a), except that the ab unit cell dimensions on the slab walls are $a = 1.05$ nm and $b = 0.6$ nm. (e) Simulated $S_1$ as a function of the applied electric field for the PVDF chains between the slab walls. Reproduced with permission.[110] Copyright 2021, Elsevier Inc.

RESPONSIVE MATERIALS

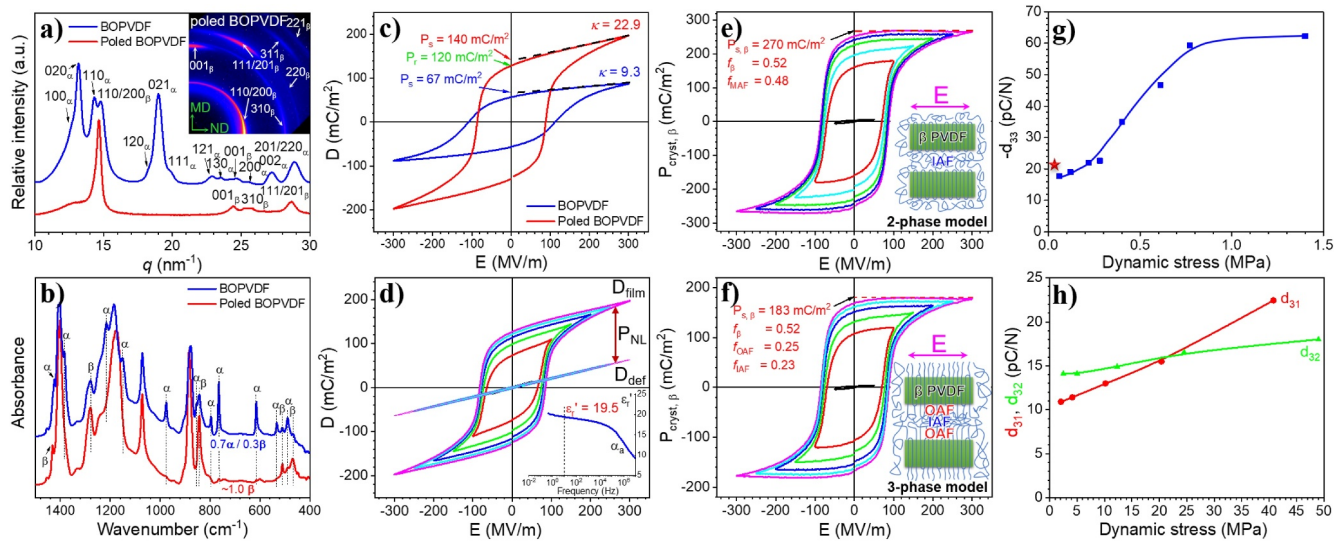

**FIGURE 9** (a) One-dimensional (1D) WAXD profiles and (b) FTIR spectra for the fresh and poled BOPVDF films at room temperature. (c) Comparison of the bipolar D-E loops for the fresh and poled BOPVDF films at 300 MV/m. (d) Progressive bipolar D-E loops for the poled BOPVDF under different poling electric fields at room temperature. The inset shows the frequency-scan real part of the relative permittivity ($\varepsilon_r'$) at 25°C. The extracted linear D-E loops from the deformational polarization ($D_{def}$) of the poled BOPVDF film are also shown. After subtracting the $D_{def}$ loop from the bipolar D-E loops for the poled BOPVDF film, nonlinear P-E loops are obtained for (e) the two-phase and (f) the three-phase models. The inset two-phase model in (e) contains the $\beta$ lamellar crystals and the isotropic amorphous fraction (IAF). The inset three-phase model in (f) contains the $\beta$ crystals, the IAF, and the oriented amorphous fraction (OAF). Using direct piezoelectric measurements, various piezoelectric coefficients are determined: (g) $d_{33}$ and (h) $d_{31}$, $d_{32}$ as a function of the dynamic stress for highly poled BOPVDF films. The red star in (g) indicates the $d_{33}$ value measured by a $d_{33}$ piezo meter with a static force of 2.5 N. Reproduced with permission.[111] Copyright 2021, Springer Nature.

In addition, we also simulated the converse piezoelectric effect: $S_1 = d_{31}E_3$, using the same strategy (Figure 8d).[110] The only difference is that we change the $ab$ unit cell dimensions on both walls to $a = 1.05$ nm and $b = 0.6$ nm. Although the equilibration time becomes longer, it is still manageable. Different electric fields are applied in the $y$-direction until an equilibrium strain is established along the $x$-direction. The electroactuation under different applied fields is shown in Figure 8e. As we can see, the $S_1$ increases with increasing the applied electric field with a slight deviation from linearity. From the slope of up to 20 MV/m, we can obtain a converse $d_{31}$ as high as 600 pm/V. This experimental prediction was later validated in a recent report.[114] Compared with the limiting $d_{31}$ value of 144.9 pC/N calculated by Tashiro and Tadokoro (Table 1),[98] our simulated $d_{31}$ value is much higher. This is most likely because of the ideal OAF chains between neighboring poled $\beta$ crystals in our simulation. Such a high limiting $d_{31}$ value stimulates us to seek further improvement in the experimental piezoelectric performance of FE polymers.

## 6 | DETERMINATION OF THE OAF CONTENT IN PIEZOELECTRIC POLYMERS

Stimulated by these computational results, we proceeded to an experimental study on a biaxially oriented PVDF (BOPVDF) film.[91,111] As revealed by the WAXD result in Figure 9a, the original film contains both $\alpha$ and $\beta$ crystals with an overall crystallinity around 0.52 (determined by DSC). From FTIR in Figure 9b, the $\alpha$ and $\beta$ contents are determined to be 0.70 and 0.30,[115] respectively, following the method in a literature report.[116] After repeated unipolar poling at 650 MV/m for at least 50 times (10 Hz), all oriented $\alpha$ crystals transform into the oriented $\beta$ phase; see 1D and 2D WAXD results in Figure 9a and FTIR in Figure 9b. The D-E loops (10 Hz) for fresh and poled BOPVDF films are shown in Figure 9c. Intriguingly, the $P_s$ increases from 67 mC/m² for the fresh BOPVDF to 140 mC/m² for the highly poled BOPVDF, which is the highest record reported in the literature. The in situ remanent polarization ($P_r$) is 120 mC/m². The apparent dielectric constant in the deformational part of the D-E loops [$\kappa = \partial D/\partial(\varepsilon_r_0 E)$] increases from 9.3 for the fresh film to 22.9 for the poled film. The inset of Figure 9d shows the frequency-scan broadband dielectric spectrum (BDS) of the poled BOPVDF at RT. At 10 Hz, the dielectric constant is 19.5, which is similar to the $\kappa$ of 22.9 at high electric fields. The linear deformational polarization loops ($D_{def}$) are determined for the D-E loops of the poled BOPVDF film (Figure 9d), following our previous report.[117] After subtracting the $D_{def}$ loops, the nonlinear $P_{NL}$-E loops are obtained, which can be used to extract the $P_s$ of the neat $\beta$ crystal ($P_{s,\beta}$). If the two-phase model is assumed, the $P_{s,\beta}$ is calculated to be 270 mC/m² (Figure 9e). This value seems impractical because the theoretical $P_{s,\beta}$ is calculated to be 185 mC/m², using DFT.[118–120] Instead, we have to assume a three-phase model with crystals, OAF, and IAF in the sample. Because of the same orientation of OAF chains as that in the crystals, they must also participate in the ferroelectric switching. To fit the theoretical $P_{s,\beta}$, the $x_{OAF}$ should be 0.25. Thus, the $x_{IAF}$ is 0.23. Intriguingly, the $-d_{33}$ of the poled BOPVDF film can reach 62 pC/N when the applied dynamic stress is above 0.7 MPa (Figure 9g). This value is about twice of that of the conventional $d_{31}$ value around 30 pC/N reported in the literature.[3,30,31] Due to the biaxial orientation, $d_{31}$ and $d_{32}$ are relatively low, only around 15 pC/N at 20 MPa. These values are similar to those reported before for biaxially oriented PVDF films.[67] If our goal is to obtain a high $d_{31}$, then the PVDF film should be uniaxially stretched.

There are several methods to determine the $x_{OAF}$ in semicrystalline ferroelectric polymers. The first method is the one described in Figure 9. Basically, we need to know the theoretical $P_s$ values of the neat FE crystals, which are not readily available for many ferroelectric polymers. Therefore, this method is relatively limited. The second method is to use TMDSC to determine the $x_{RAF}$. The RAF for PVDF has been rarely reported, except for a fast DSC study[121] and a TMDSC study.[122] Recently, we used TMDSC to calculate the $x_{RAF}$ in a PVDF homopolymer at its $T_g$ of $-45°C$: $x_{RAF} = 0.33$.[112] The crystallinity $x_c$ was 0.59 and the $x_{MAF}$ was 0.08. However, $x_{RAF}$ may not be the same as $x_{OAF}$, as shown in Figure 4d. The third method uses the full-pattern refinement to calculate the 2D WAXD pattern of the neat crystalline phase for uniaxially oriented polymers.[76–78] After subtraction of the 2D crystalline and IAF patterns from the experimental 2D fiber pattern, the 2D OAF pattern and thus $x_{OAF}$ can be obtained. However, the full-pattern refinement analysis is difficult for random copolymers, such as P(VDF-TrFE).

The fourth method also uses the 2D WAXD pattern analysis, and an example is shown in Figure 10.[86] A 2D WAXD pattern is shown for a melt-quenched and uniaxially stretched (draw ratio = 500%) P(VDF-TrFE) 75/25 mol.% (coP-75/25QS) film in Figure 10a. From the areas without crystal reflections (dashed line circles), two amorphous halos for the IAF are obtained: $IAF_1$ at 12.50 nm$^{-1}$ (0.502 nm) and $IAF_2$ at 27.86 nm$^{-1}$ (0.226 nm) (Figure 10b). $IAF_1$ is attributed to the average interchain distance and $IAF_2$ is attributed to the intrachain (or interbond) distance. After the subtraction of $IAF_1 + IAF_2$ from the experimental WAXD curve, the subtracted scattering curve represents the summation of FE crystal (K) reflections [(110/200), (001), and (111/201)] (Figure 10c), which can be fitted using the peak-fitting software. Here, we only take care of sharp crystal reflections from large primary crystals (PCs), and extremely poor secondary (or mesomorphic) crystals (SCs) are not counted. After the subtraction of the crystal reflections, the leftover curve is from the OAF, which includes poor SCs (Figure 10d). Figure 10e shows the deconvoluted experimental PVDF curve with IAF, K, and OAF deconvolution. Their contents are therefore calculated: $x_c = 0.25$,

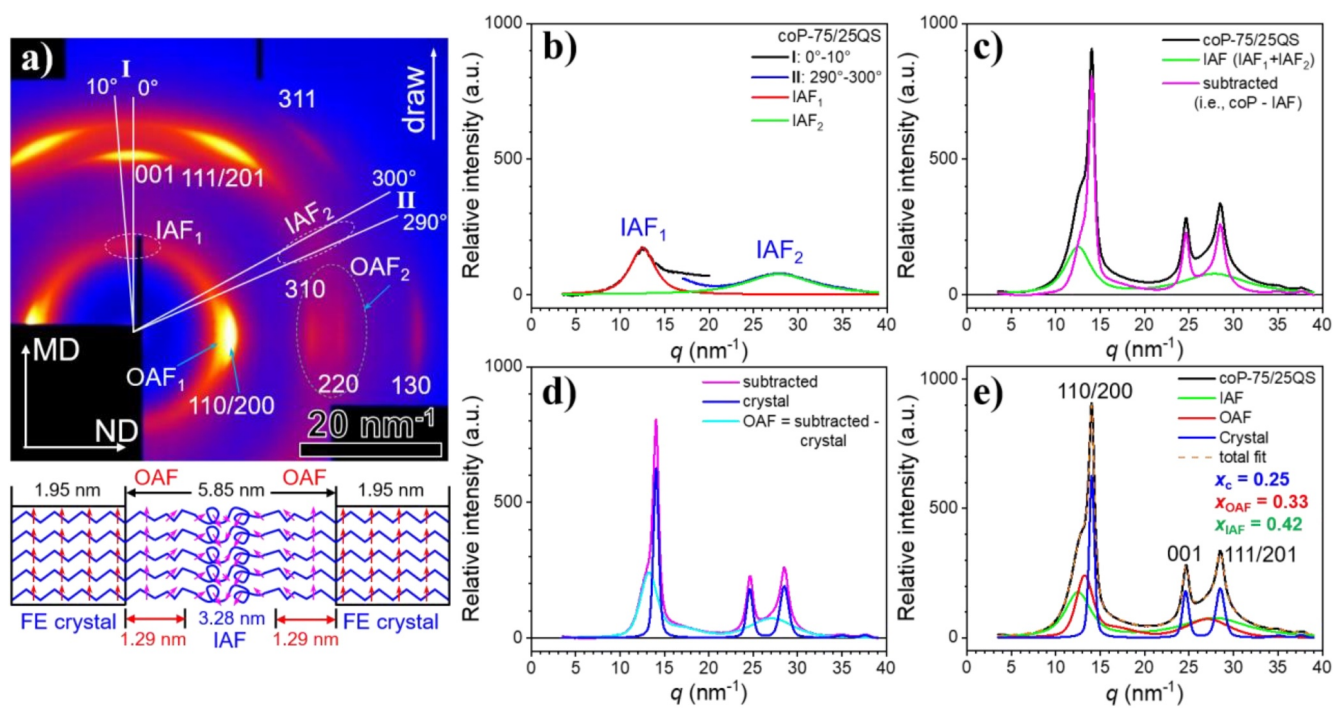

**FIGURE 10** (a) 2D WAXD pattern of the P(VDF-TrFE) 75/25 mol.% quenched and stretched (coP-75/25QS, stretching ratio = 500%) film at room temperature. The X-ray is along the transverse direction (TD) with the machine direction (MD) vertical and the normal direction (ND) horizontal. (b) Fitted $IAF_1$ [from the integration of region I (0–10°) in (a)] and $IAF_2$ scatterings [from the integration of region II (290–300°) in (a)]. (c) Subtracted WAXD curve (magenta) for crystal diffraction and OAF scattering after the sample intensity subtracts the IAF scattering (i.e., $IAF_1 + IAF_2$). (d) Subtraction of the crystalline reflections from the subtracted curve in (c) yields the OAF scattering curve (cyan). (e) The final fitting of the sample WAXD curve using crystal diffraction (blue), OAF scattering (red), and IAF scattering (green). From the integrated area, their compositions are calculated to be: $x_c = 0.25$, $x_{OAF} = 0.33$, and $x_{IAF} = 0.42$. The inset in (a) shows the schematic representation of OAF and IAF sandwiched by two crystalline lamellae. Reproduced with permission.[86] Copyright 2020, American Chemical Society.

$x_{OAF} = 0.33$, and $x_{IAF} = 0.42$. From small-angle X-ray scattering (SAXS) data, the overall lamellar thickness is 7.8 nm, as determined by correlation function analysis using the SasView software. Based on the $x_c$, $x_{OAF}$, and $x_{IAF}$ values, the crystal, ½OAF, and IAF thicknesses are calculated to be 1.95, 1.29, and 3.28 nm, respectively (see the inset of Figure 10a). Here, we assume the exclusive interlamellar OAF/IAF model for the layer thickness estimation. If there are inter-fibrillar IAF and OAF, the interlamellar OAF and IAF layer thicknesses cannot be accurately determined. However, the calculated OAF and IAF layer thicknesses should represent the upper limit values.

## 7 | RELAXOR-LIKE SCs IN OAF (SC$_{OAF}$) AND ECCs TO FURTHER ENHANCE PIEZOELECTRICITY

Using the above method, the $x_{OAF}$ can be determined from 2D fiber patterns of various P(VDF-TrFE) random copolymers with the VDF content ranging from 50 to 65 mol. %.[110,123] The samples are denoted as "coP-xx/yy", where xx is the VDF percentage and yy is the TrFE percentage. Different processing conditions were used: melt-quenched (Q) from hot-pressing, stretched (S, draw ratio ~500%), annealed (A) at 130°C for at least 12 h, and electrically poled (P) at 100 MV/m (50 MV/m DC + 50 MV/m AC at 1 Hz) and at RT. Two sets of combinations were used to make the piezoelectric films: QSP and QSAP. For example, 2D SAXS/WAXD patterns for coP-52/48QSP, coP-52/48QSAP, and coP-65/35QSAP are shown in Figure 11a,b,d,e,g,h, respectively. Four-point butterfly patterns were seen in 2D SAXS, indicating lamella tilting with respect to the stretching direction. From 2D WAXD patterns, the $x_c$, $x_{OAF}$, and $x_{IAF}$ values were determined using the fourth method described in Figure 10. For coP-52/48QSP, the overall lamellar thickness was 15.8 nm. Based on the $x_c$, $x_{OAF}$, and $x_{IAF}$ values and their densities, the crystal, ½OAF, and IAF thicknesses were obtained: 3.7, 2.2, and 7.8 nm, respectively (Figure 11c). For coP-52/48QSAP, the overall lamellar thickness was 41.7 nm, indicating that thermal annealing at 130°C (i.e., above the Curie temperature, $T_C$, around 65°C) led to an ECC structure

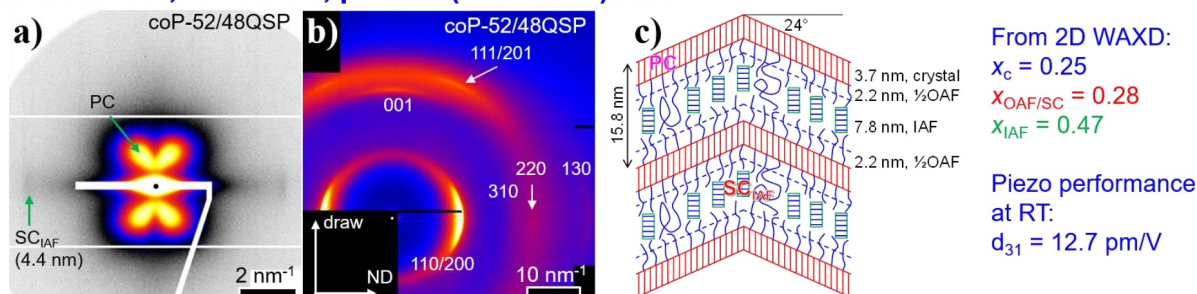
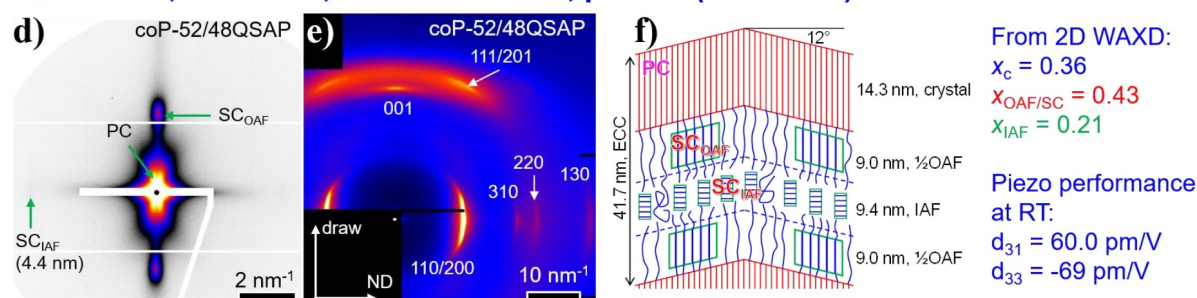
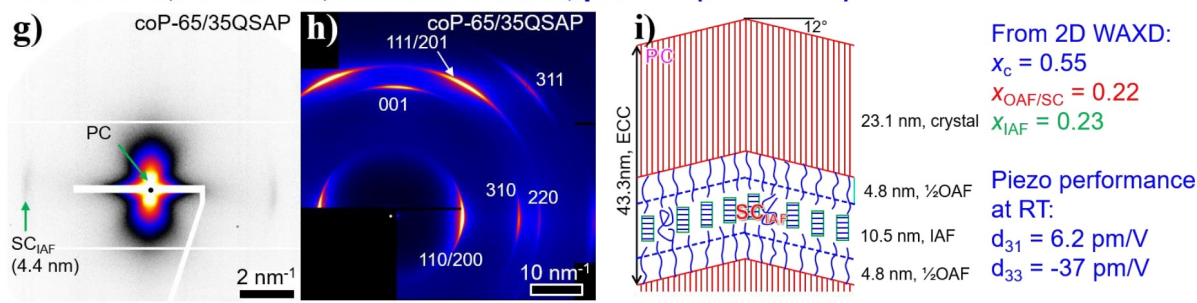

**FIGURE 11** (a, d, and g) 2D SAXS and (b, e, and h) 2D WAXD patterns of (a and b) coP-52/48QSP, (d and e) coP-52/48QSAP, and (g and h) coP-65/35QSAP films. Proposed semicrystalline structures for (c) coP-52/48QSP, (f) coP-52/48QSAP, and (i) coP-65/35QSAP films with calculated crystallinity ($x_c$), OAF and IAF contents ($x_{OAF}$ and $x_{IAF}$), and piezoelectric performance. Reproduced with permission.[110] Copyright 2021, Elsevier Inc.

in the sample. It is wellknown that thermal annealing of P(VDF-TrFE) random copolymers above the $T_C$ will lead to ECCs,[124–126] and this is attributed to the enhanced chain-sliding motion in the PE crystalline phase.[127,128] Based on the $x_c$, $x_{OAF}$, and $x_{IAF}$ values and their densities, the crystal, ½OAF, and IAF thicknesses were obtained: 14.3, 9.0, and 9.4 nm, respectively (Figure 11f). For coP-65/35QSAP, the overall lamellar thickness was 43.3 nm, indicating again the ECC structure after thermal annealing at 130°C. Based on the $x_c$, $x_{OAF}$, and $x_{IAF}$ values and their densities, the crystal, ½OAF, and IAF thicknesses were obtained: 23.1, 4.8, and 10.5 nm, respectively (Figure 11i).

The $S_1$-$E$ loops at 10 MV/m and 1 Hz were used to determine the converse $d_{31}$ for various QSP (Figure 12a) and QSAP copolymer films (Figure 12b). In general, all QSP samples exhibited low $d_{31}$, whereas QSAP samples exhibited high $d_{31}$ except for coP-65/35QSAP, which had the lowest $x_{OAF}$ and the highest $x_c$. The high $d_{31}$ for QSAP copolymers with a low VDF content (<65 mol.%) could be attributed to the formation of relaxor-like $SC_{OAF}$ after electric poling; see the 2D SAXS pattern in Figure 11d and the corresponding schematic representation in Figure 11f. Note that the thick ½OAF layers (ca. 9.0 nm) favored the formation of $SC_{OAF}$, which exhibited an obvious melting shoulder at 38°C before the $T_C$ peak at 70°C in temperature-scan BDS (Figure 12c). Because the maximum temperature of this shoulder peak was slightly frequency dependent (which is a typical behavior for RFE ceramics[129,130] and polymers[3,131,132]), we call it relaxor-like $SC_{OAF}$. Such a low-temperature melting peak is often observed for PVDF homopolymers and P(VDF-TrFE) copolymers, and it has puzzled researchers for many years.[102] In some reports, it is even attributed to a second $T_g$ from the constrain amorphous phase or RAF.[102,133,134] Through this study, the origin of this low temperature transition can be clearly attributed to the melting of the relaxor-like $SC_{OAF}$. It must be the easy conformation transformation (i.e., electrostriction motion) in the $SC_{OAF}$ (and OAF) that enhances the piezoelectric performance. On the contrary, the low $d_{31}$ for all QSP samples could be attributed to the small ½OAF thickness of only ~2–3 nm (e.g., see coP-52/48QSP in Figure 11c), which was too thin to grow $SC_{OAF}$. For coP-65/35QSAP, the $x_{OAF}$ became low (0.22) because most OAF chains had crystallized into the thick lamellar crystals during thermal annealing in the PE phase at 130°C (i.e., above its $T_C$ of 108°C). As a result, the ½OAF thickness was only 4.8 nm (Figure 11i), which was also too thin to grow $SC_{OAF}$. No shoulder peak was observed in the temperature-scan BDS for coP-65/35QSAP.[110] Consequently, it exhibited a very low $d_{31}$.

Upon heating, the QSAP samples of P(VDF-TrFE) with a low VDF content (<65 mol.%) could exhibit an increased $d_{31}$. An example is shown for coP-55/45QSAP in Figure 12d–f. As the temperature increased, the D-E loops became slimmer with a double hysteresis shape (Figure 12d). As a result, the $P_r$ gradually decreased. However, the $d_{31}$ increased to 77 pm/V at 55°C (Figure 12e,f), after which the piezoelectricity disappeared at 70°C (i.e., close to its $T_C$ at 75°C). The electromechanical coupling factor ($k_{31}$) reached a maximum value of 12.5% around 50°C. The maximum $d_{31}$ at 55°C was attributed to the melting of relaxor-like $SC_{OAF}$ at 57°C observed in the temperature-scan BDS (Figure 12c).[110] As seen in Figure 12e, the $S_1$-$E$ loop at 65°

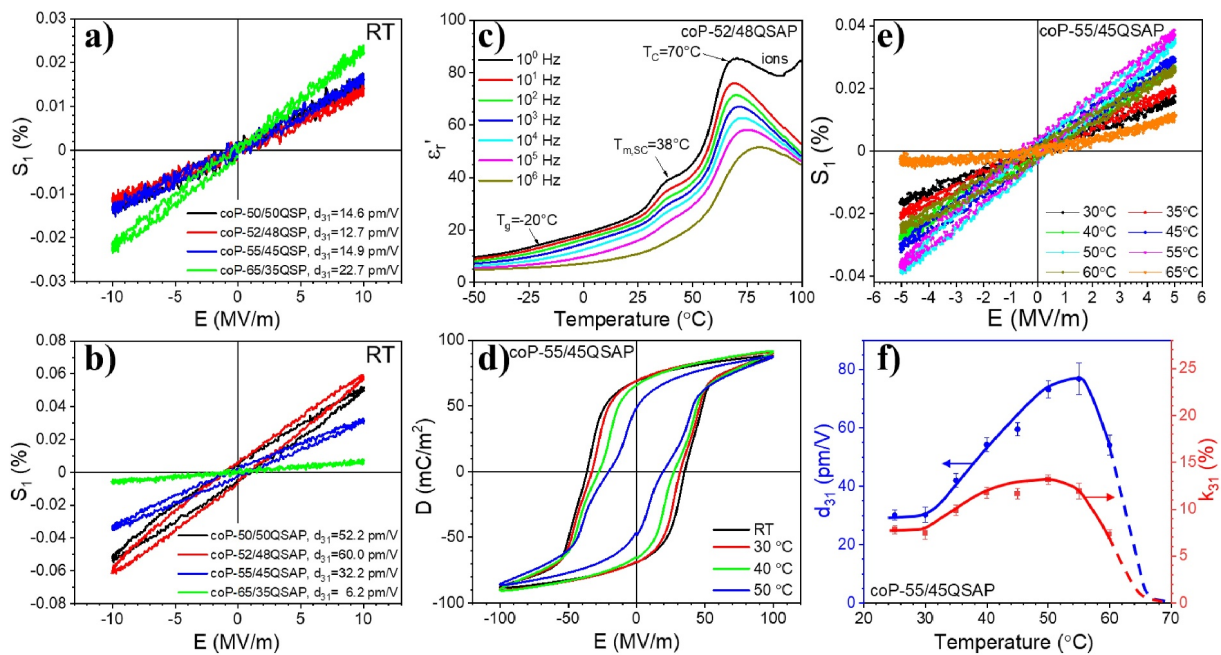

**FIGURE 12** Determination of the converse $d_{31}$ using bipolar $S_1$-$E$ loops for various (a) QSP and (b) QSAP P(VDF-TrFE) copolymer films at room temperature. (c) Temperature-scan $\varepsilon_r'$ BDS of for coP-52/48QSAP under different frequencies. (d) Bipolar D-E loops at different temperatures, (e) bipolar $S_1$-$E$ loop at different temperatures, and (f) temperature-dependent $d_{31}$ and $k_{31}$ for the coP-55/45QSAP film. Reproduced with permission.[110] Copyright 2021, Elsevier Inc.

C became a combination of piezoelectric and electrostrictive behavior, and the $d_{31}$ could not be unambiguously determined.

During the review process, a question is asked about the melting temperature ($T_m$) of P(VDF-TrFE) random copolymers. It is known that the thicker the crystalline lamellar thickness, the higher the $T_m$ for semicrystalline homopolymers. However, this is usually not the case for isomorphic P(VDF-TrFE) crystals, as observed in previous reports.[110,135] In addition to the crystalline lamellar thickness, the defects (i.e., TrFE) inside the crystalline region can decrease the $T_m$ (and the heat of fusion as well). As the crystalline lamellae thicken, more TrFE defects are included in the crystalline region; therefore, the $T_m$ should decrease. Due to the interplay between these two opposing effects, the experimentally observed $T_m$ only increases slightly for P(VDF-TrFE) upon thermal annealing above the $T_C$.

## 8 | FORMATION OF $SC_{OAF}$ BY HIGH-POWER ULTRASONICATION AND THE EFFECT OF HEAD-TO-HEAD AND TAIL-TO-TAIL (HHTT) DEFECTS ON PIEZOELECTRICITY IN PVDF HOMOPOLYMERS

Although relatively high $d_{31}$ can be obtained for P(VDF-TrFE) copolymers with a low VDF content, their disadvantages are obvious. First, the low $T_C$ of these copolymers (<75°C) prevent high temperature applications. Second, the price of P(VDF-TrFE) copolymers is astonishingly high ($10 k/kg). Therefore, it is still highly desired to enhance the piezoelectric performance of PVDF homopolymers, which has a high thermal stability and a low price of ~$15/kg. From the above study, we have learned that relaxor-like $SC_{OAF}$ in the ECC structure is important to enhance piezoelectricity via electrostriction. However, it is not easy to achieve ECCs for PVDF homopolymers without high-pressure crystallization. From the temperature-pressure phase diagram of PVDF, the triple point (where the isotropic melt, FE $\beta$ crystal, and PE crystal coexist) is located around 300°C and 300 MPa.[127,136] Supposedly, only above 300 MPa and 300°C (note that PVDF can autonomously degrade and release HF above 300°C), the PE phase can be accessed by heating above the $T_C$. This is the reason why under ambient pressure, PVDF homopolymer does not show any $T_C$ below its $T_m$ around 175°C.[137] By copolymerizing TrFE into PVDF, the triple point gradually decreases to lower temperatures and pressures.[138] Finally, when the TrFE content is above 18 mol.%, the triple point reaches the ambient pressure. This is why P(VDF-TrFE) random copolymers with VDF content below 82 mol.% can automatically exhibit the FE phase with a $T_C$.[89,138,139] Ohigashi and coworkers used a pressure cell to grow PVDF ECCs with a pressure quench procedure.[140] After high temperature electric poling (130 MV/m at 120°C) to achieve 100% $\beta$ ECCs with a large $P_{r0}$, the $\beta$-ECC PVDF sample exhibited a high electromechanical coupling factor $k_{33}$ of 0.27, which persisted up to 180°C with an only 20% loss of performance.[140] We consider that the $SC_{OAF}$ in the ECC structure of these ECC PVDF samples must have played an important role, similar to the case of QSAP P(VDF-TrFE) copolymers with a low VDF content (<65 mol.%).

However, it is difficult to mass produce PVDF films using the high-pressure crystallization method. It is highly desired to increase the piezoelectric performance of commonly processed PVDF samples. Recently, we discovered a high-power ultrasonication method to obtain the relaxor-like $SC_{OAF}$ in conventionally processed PVDF films.[141,142] The schematic principle is presented in Figure 13. The high-power ultrasonic energy can break the tips of FE PCs to obtain the relaxor-like $SC_{OAF}$; see Figure 13a,b). These relaxor-like $SC_{OAF}$ crystals can enhance the piezoelectricity of PVDF homopolymers (see Figure 13b,c).

In this work, we studied two PVDF homopolymers with the HHTT defect content being 4.3% and 5.9%, respectively. The TT content was determined from proton nuclear magnetic resonance ($^1$H NMR) spectroscopy (Figure 14a), and the HH content was determined from $^{19}$F NMR (Figure 14b). The PVDF with 4.3% HHTT defects exhibited a high $T_m$ at 176°C (it is called HMT PVDF) and the PVDF with 5.9% HHTT defects exhibited a low $T_m$ at 162°C (it is called LMT PVDF). After melt-quench from hot-pressing, uniaxial stretching (S, ~500% draw ratio), and unipolar poling (P, 200 MV/m DC + 200 MV/m AC at 1 Hz), the samples were subjected to high-power (300 W) ultrasonic processing (U). Here, we use the LMT PVDF sample as an example to demonstrate the effect of ultrasonication. As shown in Figure 15a–c, the ultrasonication time played an important role in the piezoelectric performance. Before 20 min, the $d_{31}$ increased with increasing ultrasonication time (Figure 15b). Beyond 20 min, the film sample was partly damaged by the high-energy ultrasonication and exhibited a reduced $d_{31}$. Therefore, the optimal ultrasonication time was around 20 min. When a high unipolar field (100 MV/m) was applied, the effect of ultrasonication was clearly seen in Figure 15c. Moreover, without ultrasonication, a hard piezoelectric behavior was seen with a low $d_{31}$. As the ultrasonication time increased, the sample transformed into an increasingly soft piezoelectric with broadened $S_1$-$E$ loops. This is similar to the hard-to-soft piezoelectric transition for ceramic piezoelectrics reported in the literature,[28,29] where soft piezoelectrics can exhibit significantly higher piezoelectric coefficients (e.g., $d_{33}$ ~ 2000 pC/N for lead manganese titanate–lead titanate[29]).

With a fixed ultrasonication time of 20 min, two samples were studied using 2D SAXS and WAXD: HMT-SPU and LMT-SPU PVDF films (Figure 14e–h). From the SAXS curves, the lamellar spacings were 9.75 and 6.08 nm for HMT-SPU and LMT-SPU, respectively (Figure 14g). From the 2D WAXD patterns in Figure 14e,f, the $x_c$, $x_{OAF/SC}$ (note that it is not possible to differentiate scatterings from OAF and poor $SC_{OAF}$) and $x_{IAF}$ were determined (Figure 14h).

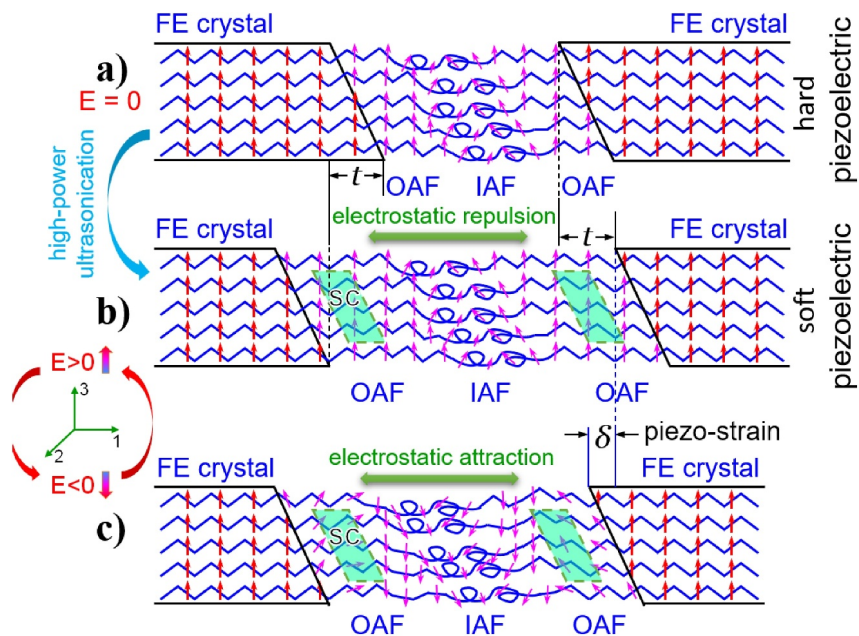

**FIGURE 13** Schematic formation of $SC_{OAF}$ by high-power ultrasonication and the converse piezoelectricity in ferroelectric (FE) PVDF. (a) PVDF-SP at $E = 0$ and PVDF-SPUx at (b) $E > 0$ and (c) $E < 0$. Red and magenta arrows are the VDF dipoles in the poled $\beta$ crystals and the amorphous phase (OAF + IAF), respectively. The green parallelograms in the OAF are SCs. Reproduced with permission.[141] Copyright 2022, the Royal Society of Chemistry.

The HMT-SPU film had a higher $x_c$ but a lower $x_{OAF/SC}$, whereas the LMT-SPU film had a lower $x_c$ and a higher $x_{OAF/SC}$. The $SC_{OAF}$ was detected by DSC and BDS to have a low $T_m$ around 60°C, and its content was only about 2.5%. Since LMT-SPU had a higher $x_{OFA/SC}$, we expect that it should have better piezoelectric performance than HMT-SPU.

In Figure 15d, we compare the high-field unipolar piezoelectric performance between HMT-SP/LMT-SP (hard piezoelectric) and HMT-SPU/LMT-SPU samples (soft piezoelectric). Indeed, LMT-SP and LMT-SPU had higher piezoelectric performance than did HMT-SP and HMT-SPU. Temperature-dependent $d_{31}$ and $k_{31}$ for various PVDF films are shown in Figure 15e,f, respectively. Again, LMT-SP and LMT-SPU exhibited higher $d_{31}$ than HMT-SP and HMT-SPU. For LMT-SPU, the maximum $d_{31}$ reached 76.2 pm/V at 65°C, and its thermal stability reached 100°C. For HMT-SPU, the maximum $d_{31}$ reached 68 pm/V at 65°C, and its thermal stability reached 110°C. After the first heating and melting of the relaxor-like $SC_{OAF}$ crystals, both HMT-SPU (Figure 15g) and LMT-SPU (Figure 15h) exhibited lower $d_{31}$ values. HMT-SPU had a $d_{31}$ of ~58 pm/V for temperature above 60°C. LMT-SPU had a $d_{31}$ of ~64 pm/V for temperature above 50°C. Below these temperatures, $d_{31}$ decreased for both HMT-SPU and LMT-SPU because of the crystallization of some SCs between the PC lamellae. Note that these SCs formed by non-isothermal crystallization should be similar to the $SC_{OAF}$ obtained by high-power ultrasonication; however, their amount must be significantly lower. Again, they are responsible for the mid-temperature transitions in PVDF and P(VDF-TrFE) polymers discussed before.[102]

## 9 | CONCLUSIONS AND FUTURE OUTLOOK

In this perspective, our intention was to unravel the fundamental physics of piezoelectricity in solid films of FE polymers, such as PVDF and P(VDF-TrFE). The major contribution to polymer piezoelectricity was first attributed to the high Poisson's ratio ($v_{31} = 0.6$–$0.7$) around 1980 and more recently to the OAF in semicrystalline polymers. MD simulations not only demonstrate the OAF mechanism (i.e., conformation transformation) for direct and converse piezoelectricity, but also predict the limiting $d_{31}$ value to be as high as 600 pm/V. This prediction encourages us to continue experimental studies to enhance the piezoelectric performance of FE polymers via electrostriction in OAF. In addition to OAF, the relaxor-like $SC_{OAF}$ are found to be more influential for the enhanced piezoelectricity. Ideally, the ECC structure is preferred for the growth of relaxor-like $SC_{OAF}$, and it can be easily achieved for P(VDF-TrFE) with a VDF content around 50 mol.%. However, ECCs are difficult to achieve for PVDF homopolymers, which have a higher thermal stability. High-power ultrasonication is used to induce the $SC_{OAF}$ in PVDF homopolymers without the ECC structure. Furthermore, it is found that a higher level of the HHTT defects can promote the $SC_{OAF}$ formation to increase $d_{31}$.

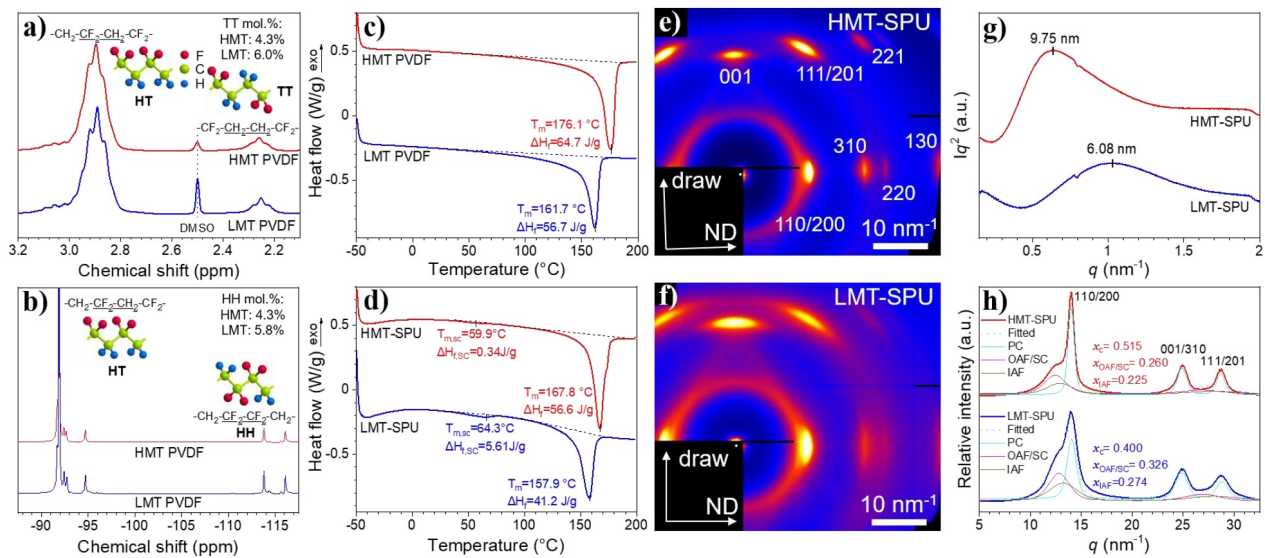

**FIGURE 14** (a) $^1$H and (b) $^{19}$F NMR spectra of HMT and LMT PVDF in d$_6$-DMSO. (c) Second heating DSC curves after removing the prior thermal history for HMT and LMT PVDF resins. (d) First heating curves for uniaxially stretched, poled, and ultrasonicated (SPU) HMT and LMT films (HMT-SPU and LMT-SPU). 2D WAXD patterns for uniaxially stretched (e) HMT-SPU and (f) LMT-SPU PVDF films at room temperature. The X-ray intensity is in a logarithmic scale. (g) 1D WAXD results from (e) and (f). Peak deconvolution is performed using the Peakfit software to determine (h) PC, OAF/SC, and IAF contents. Reproduced with permission.[142] Copyright 2023, Elsevier Inc.

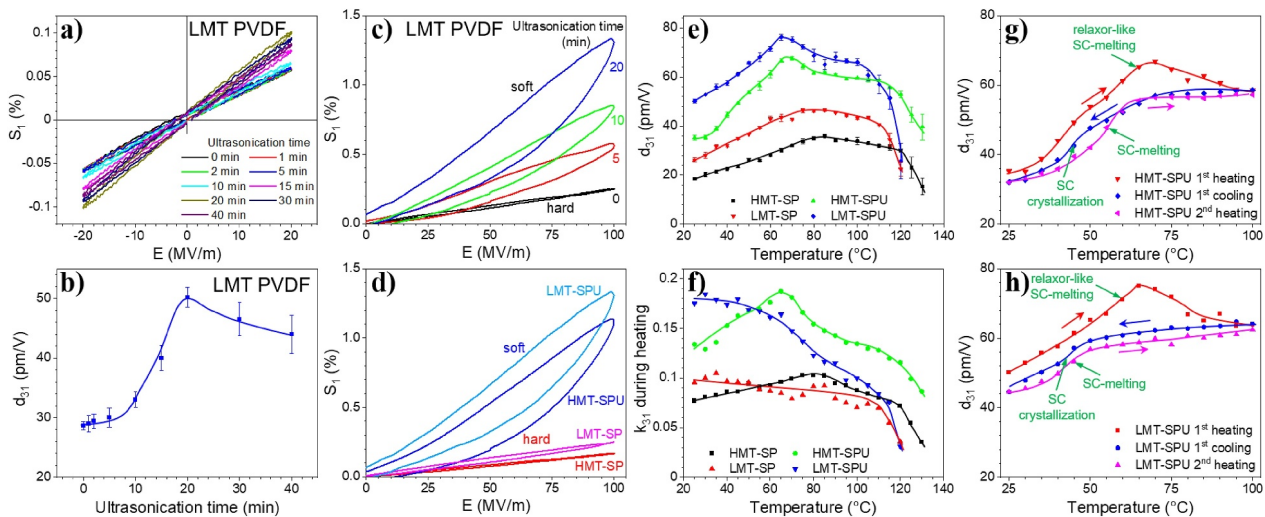

**FIGURE 15** (a) Low-field $S_1$-$E$ loops for different PVDF-SPUx films (x indicates different lengths of ultrasonication time in minutes). (b) Calculated $d_{31}$ values as a function of the ultrasonication time. (c) High-field unipolar $S_1$-$E$ loops for different PVDF-SPUx films. (a–c) Reproduced with permission.[132] Copyright 2022, Royal Society of Chemistry. (d) Unipolar $S_1$-$E$ loops of the HMT-SPU and LMT-SPU PVDF films. Calculated (e) converse $d_{31}$ and (f) $k_{31}$ for various PVDF samples: HMT-SP, LMT-SP, HMT-SPU, and LMT-SPU. Converse $d_{31}$ during the first heating, the first cooling, and the second heating processes for (g) HMT-SPU and (h) LMT-SPU PVDF films. (d–h) Reproduced with permission.[102] Copyright 2023, Elsevier, Inc.

The findings in this perspective just represent the tip of an iceberg. There are still needs for further exploration to enhance piezoelectricity of polymers to be comparable to that of their ceramic counterparts:

1) The direct link between the high Poisson's ratio and the OAF in uniaxially stretched semicrystalline polymers needs to be clearly established. Both simulation and experimental studies should be conducted. For example, MD simulations of the composite model in Figure 5 can be carried out for situations with and without the OAF. Experimentally, the $x_{OAF}$ should be determined for PVDF samples stretched at different draw ratios and correlated with the Poisson's ratio $v_{31}$. If a positive correlation is found between $x_{OAF}$ and $v_{31}$, it is certain that the OAF is the reason for high $v_{31}$ found in uniaxially stretched semicrystalline polymers.

2) The $SC_{OAF}$ content in ultrasonicated PVDF is still too low, at only 2.5%. It is highly desirable to further increase the $SC_{OAF}$ content to increase piezoelectricity. Based on our results in Figure 15, we should further increase the HHTT content for PVDF. An HHTT content of 5.9% is the limit for the current commercially available PVDF resins. It is desirable for industries to mass produce new (and cheap) PVDF homopolymers with even higher HHTT contents and to explore their piezoelectricity after ultrasonication.

3) As demonstrated in a RFE PVDF-based tetrapolymer, an ultrahigh $-d_{33}$ around 1000 pm/V could be achieved at a bias electric field of 40 MV/m.[114] This experimental result not only confirms our simulated limiting values of piezoelectric coefficient, but also demonstrates the electrostrictive origin for piezoelectricity. However, this PVDF-based tetrapolymer is not a genuine piezoelectric polymer, but an electrostrictive polymer. Applying a 40 MV/m bias field to realize high piezoelectricity is not practical; therefore, it is highly desired to design and synthesize a genuine piezoelectric polymer with such a high piezoelectric coefficient. With such a high piezoelectric coefficient, the electromechanical coupling factor can reach as high as 0.88,[114] which is highly desirable for use in transducers for ultrasonic imaging/therapy and energy harvesting applications.[9]

4) The electrostrictive OAF mechanism for piezoelectricity should be applicable to other FE polymers. For example, odd-numbered nylons (e.g., nylon-11) represent another family of FE polymers.[108,143] Thermodynamically, the α phase of odd-numbered nylons having an all-trans conformation and a parallel arrangement of amide dipoles in the hydrogen-bonding sheets is the most stable polar phase for piezoelectricity. However, because of the strong hydrogen-bonding sheets, the α phase is non-ferroelectric and macroscopic samples with a random domain orientation cannot be polarized to exhibit $P_{r0}$ and piezoelectricity. Instead, the mesomorphic δ phase (obtained from melt-quenching) with random and weak hydrogen-bonding is needed to achieve ferroelectricity. After electric poling of the δ-phase samples, a sizable $P_{r0}$ is obtained, and piezoelectricity results. However, the reported $d_{31}$ for poled odd-numbered nylon films is rather low at RT, often below 5 pC/N.[144] This is attributed to the frozen OAF (i.e., RAF) below the $T_g$ around −45°C. Only when the temperature is above 120°C to promote the dipole mobility in the OAF, the $d_{31}$ increases to 15–20 pC/N for uniaxially stretched samples (draw ratio >300%).[145] However, extended annealing at >120°C will decrease the piezoelectricity for nylon-11 because the δ phase gradually transforms into the α′ phase, which has stronger hydrogen-bonding, decreased unit cell dimensions, and thus reduced dipole mobility in the OAF. To further enhance the piezoelectric performance for odd-numbered nylons, it is highly desirable to increase the OAF mobility by decreasing the $T_g$ to a value far below RT (e.g., by using plasticizers[146]). New nylon copolymers with methylated amide groups should be systematically studied.[147]

5) The electrostrictive OAF mechanism for piezoelectricity should also be applicable for solid films of piezoelectric biopolymers, which are preferable to nondegradable PVDF-based fluoropolymers, as their biodegradability leads to reduced environmental impacts. As we mentioned above, however, the current challenge for piezoelectric biopolymers is their low piezoelectric performance. There is a great need for the design and synthesis of new biopolymers, such as PLLA and PHA copolymers, with high dipole mobility in the OAF to enhance their piezoelectricity.

6) The electrostrictive OAF mechanism can also help us design better piezoelectric fiber mats, although the major working mechanism is the dimensional model. With enhanced piezoelectricity in the solid fibers, we envision that the porous fiber mats should have improved piezoelectric performance. Currently, electrospinning is primarily used to achieve self-polarized fiber mats. However, electrospinning is difficult to scale up for mass production. Viable post-electric poling processes are necessary to achieve $P_{r0}$ and piezoelectricity for FE polymer fiber mats.

7) Once the desired piezoelectric performance is achieved for the above-mentioned piezoelectric polymers, it will be necessary to implement viable processing methods for the mass production of piezoelectric films and fiber mats in the areas of sensors,[148] actuators,[149] and energy harvesters.[150,151] Note that solution-processing methods use polar solvents such as N,N-dimethylformamide,[152] they are not environmentally benign, and are more expensive, as compared to melt-processing. For PVDF and PLLA/PHA, melt-casting followed by uniaxial stretching can be used to mass produce piezoelectric films (note that in-line corona poling is needed to polarize the PVDF films). Although some companies such as PolyK and Measurement Specialties (sold to TE Connectivity) are practicing the melt-processing method, process conditions need to be further optimized to achieve higher piezoelectric performance. Meanwhile, device engineering should be conducted to integrate various piezoelectric films and fiber mats into the wealth of devices, such as ultrasonic imaging probes, haptic sensors, soft robotics, and energy harvesting units, where they would add significant value.[9]

## AUTHOR CONTRIBUTIONS

**Guanchun Rui**: Conceptualization (lead); data curation (lead); methodology (lead); resources (lead); validation (lead); writing – review & editing (lead). **Elshad Allahyarov**: Data curation (equal); formal analysis (equal); methodology (equal); software (lead); validation (lead); visualization (lead). **Zhiwen Zhu**: Conceptualization (equal); data curation (equal); methodology (equal); resources (equal); validation (equal). **Yanfei Huang**: Data curation (equal); investigation (equal); methodology (equal); resources (equal); validation (equal). **Thumawadee Wongwirat**: Conceptualization (supporting); data curation (equal); formal analysis (equal); investigation (supporting); methodology (supporting). **Qin Zou**: Data curation (equal); formal analysis (equal); investigation (supporting); methodology (supporting). **Philip L. Taylor**: Project administration (lead); software (equal); supervision (lead); validation (equal); visualization (equal); writing – review & editing (equal). **Lei Zhu**: Conceptualization (lead); funding acquisition (lead); investigation (lead); project administration (lead); supervision (lead); writing – original draft (lead).

### ACKNOWLEDGMENTS
L. Zhu and P. L. Taylor acknowledge financial support from National Science Foundation, Division of Materials Research, Polymers Program (DMR-2103196). E. Allahyarov acknowledges financial support from the Ministry of Science and Higher Education of the Russian Federation (State Assignment No. 075-00270-24-00).

### CONFLICT OF INTEREST STATEMENT
The authors declare no conflict of interests.

### DATA AVAILABILITY STATEMENT
The data that support the findings of this study are available from the corresponding author upon reasonable request.

### ORCID
*Guanchun Rui*  https://orcid.org/0000-0002-4097-2205
*Lei Zhu*  https://orcid.org/0000-0001-6570-9123

## AUTHOR BIOGRAPHIES

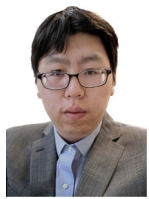

**Guanchun Rui** received his Ph.D. degree from the Department of Macromolecular Science and Engineering, Case Western Reserve University, in 2023, under the advice of Prof. Lei Zhu. He is currently a postdoctoral scientist at Arkema, Inc., and a visiting scholar at the Pennsylvania State University. His research interests include semicrystalline polymers and electroactive polymers, especially for their piezoelectric, electrostrictive, and electrocaloric applications.

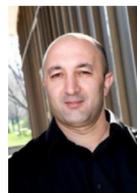

**Elshad Allahyarov** is currently a researcher at the Department of Physics of Case Western Reserve University. He also holds scientific positions at the Institute for Theoretical Physics II in Heinrich-Heine University of Dusseldorf, Germany, and in the Theoretical Department of the Institute for High Temperatures of Russian Academy of Sciences in Moscow, Russia. He received his Ph.D. (in 1994) and Doctor of Sciences (in 2005) degrees in Theoretical Physics from the Russian Academy of Sciences. His research interests include theoretical soft matter physics, demixing and phase separation in charged and dipolar systems under external loads, and capacitive energy storage in high-$\kappa$ dielectric/ferroelectric polymers and polymer nanocomposites.

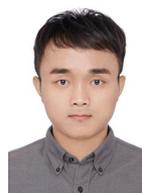

**Zhiwen Zhu** received his Ph.D. degree in Mechanical Engineering from South China University of Technology, Guangzhou, China, in 2021. From 2019 to 2021, he was a visiting student in Prof. Lei Zhu's group at Case Western Reserve University. He is currently pursuing postdoctoral research at the School of Mechanical and Automotive Engineering in South China University of Technology. His research interests focus on piezoelectric, ferroelectric, and dielectric polymers.

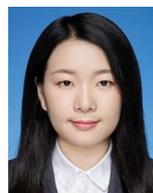

**Yanfei Huang** is currently an assistant professor of College of Materials Science and Engineering at Shenzhen University. She received her Ph.D. degree in Materials Processing from Sichuan University in 2018. From 2016 to 2018, she was a visiting student in Case School of Engineering at Case Western Reserve University. Before joining the faculty at Shenzhen University in 2020, she was a post-doctoral researcher at Tsinghua University. Her research interests include piezoelectric, ferroelectric, and dielectric polymers, and solid-state polymer electrolytes for lithium-ion batteries.

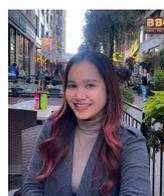

**Thumawadee Wongwirat** received her Ph.D. in polymer science from the Petroleum and Petrochemical College, Chulalongkorn University, Thailand. She joined Prof. Lei Zhu's research group in 2021 as a research associate in the Department of Macromolecular Science and Engineering, Case Western Reserve University. Her research interests include dielectric/ferroelectric polymers, thermal conductive polymers composites, and multilayer coextrusion for polymer processing.

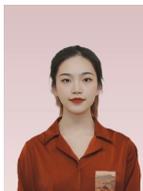

**Qin Zou** is a Ph.D. student in the Department of Macromolecular Science and Engineering at Case Western Reserve University. She received her bachelor's degree in Polymer Science and Engineering at South China University of Technology, and joined Prof. Lei Zhu's research group in 2022. Her current research interest focuses on the electrostriction of relaxor ferroelectric polymers.

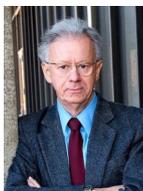

**Philip L. Taylor** is a Distinguished University Professor, and Perkins Professor of Physics and Macromolecular Science Emeritus, at Case Western Reserve University, where he has been since earning his Ph.D. from the University of Cambridge in 1962. His wide-ranging research interests in theoretical physics include the thermodynamic properties of polymers and liquid crystals as well as the transport properties of metals and the quantum hall effect.

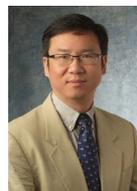

**Lei Zhu** received his Ph.D. degree in Polymer Science from University of Akron in 2000. He joined the Institute of Materials Science and Department of Chemical, Materials and Biomolecular Engineering at the University of Connecticut in 2002, as an assistant professor. In 2009, he moved to the Department of Macromolecular Science and Engineering at Case Western Reserve University as an associate professor. In 2013, he was promoted to full professor. His research interests focus on structure-property-processing relationships in high-$\kappa$ dielectric/ferroelectric polymers and polymer nanocomposites for advanced electrical and electronic applications.